\pdfoutput=1
\documentclass[acmtog, authorversion]{acmart}

\usepackage{booktabs} % For formal tables

\setcitestyle{square}

\usepackage[ruled]{algorithm2e} % For algorithms

\SetAlFnt{\small}
\SetAlCapFnt{\small}
\SetAlCapNameFnt{\small}
\SetAlCapHSkip{0pt}
\IncMargin{-\parindent}

\setcopyright{usgovmixed}

\usepackage[ruled]{algorithm2e} % For algorithms
\usepackage[colorinlistoftodos]{todonotes}

\SetAlFnt{\small}
\SetAlCapFnt{\small}
\SetAlCapNameFnt{\small}
\SetAlCapHSkip{0pt}
\IncMargin{-\parindent}

\usepackage{listings}
\usepackage{amsmath}
\usepackage{array,multirow,booktabs}
\usepackage[toc,page]{appendix}
\usepackage{color}
\definecolor{mygreen}{rgb}{0,0.6,0}
\definecolor{mygray}{rgb}{0.5,0.5,0.5}
\definecolor{mymauve}{rgb}{0.58,0,0.82}

\lstdefinestyle{CGA}{ %
	language=C, %
	basicstyle=\scriptsize, %
	frame=single, %
	title=\lstname, %
	commentstyle=\color{mygreen}, %
	stringstyle=\color{mymauve}, %
	keywordstyle=\color{blue}, %
	deletestring=[b]',%
	deletekeywords={type}, %
	otherkeywords={attr,@StartRule,NIL,%
		split,extrude,comp,entrance,zone,%
		members,count,set,goToBuilding,goToObject,stayInside,delayedRule,goToZone,wait,%
		accompany,findBuilding,findNearestBuilding,getDistanceInTime,chooseMember,%
		findObject,interact,%
		floatingSlot, floatingTask%
		@Object%
	}, %
	escapeinside={\%*}{*)}, %
	rulecolor=\color{black}, %
	keepspaces=true, %
	numbers=left, %
	numbersep=5pt, %
	numberstyle=\tiny\color{mygray}, %
	breaklines=true, %
	breakatwhitespace=false, %
	showspaces=false, %
	showstringspaces=false, %
	showtabs=false, %
	tabsize=2 %
}

\newcommand{\TODO}[3]{\todo{\textcolor{#2}{\textit{#3}}]}}

\definecolor{darkgreen}{RGB}{39, 174, 96}
\iftrue
\newcommand{\OtgerTODO}[1]{\TODO{Otger}{darkgreen}{#1}}
\newcommand{\GusTODO}[1]{\TODO{Gus}{blue}{#1}}
\newcommand{\NuriaTODO}[1]{\TODO{Nuria}{magenta}{#1}}

\fi
\newcommand{\change}[1]{{\color{black}#1}}

\hypersetup{draft}
\begin{document}
	
	% \title{A Multifrequency MAC Specially Designed for Wireless Sensor
	%   Network Applications}
	% \affiliation{%
	% \email{beranger@inria.fr}
	
	\title%[Procedural Simulation of Crowds in Semantically Augmented Virtual Cities]
	{Procedural Crowd Generation for \\ Semantically Augmented Virtual Cities}
	
	%\author[papers\_401s2]{ID: papers\_401s2}
	
	% for anonymous conference submission please enter your SUBMISSION ID
	% instead of the author's name (and leave the affiliation blank) !!
	\author%[O. Rogla \& N. Pelechano \& G. Patow]
	{O. Rogla$^{1}$,
		N. Pelechano$^{1}$
		and G. Patow$^{2}$
		% %        S. Spencer$^2$\thanks{Chairman Siggraph Publications Board}
		\\
		% % For Computer Graphics Forum: Please use the abbreviation of your first name.
		$^1$UPC, Universitat Politecnica de Catalunya , Spain\\
		$^2$UdG, Universitat de Girona, Spain\\
		%          %$^2$Institut f{\"u}r ComputerGraphik \& Wissensvisualisierung, TU Graz, Austria
		% %        $^2$ Another Department to illustrate the use in papers from authors
		% %             with different affiliations
	}
	% \thanks{This work is supported by Anonymous.
	%   Author's addresses: Anonymous.
	%   }
	\renewcommand\shortauthors{O. Rogla \& N. Pelechano \& G. Patow}
	
	\begin{teaserfigure}
		%\teaser{
		\centering
		\includegraphics[width=1\linewidth]{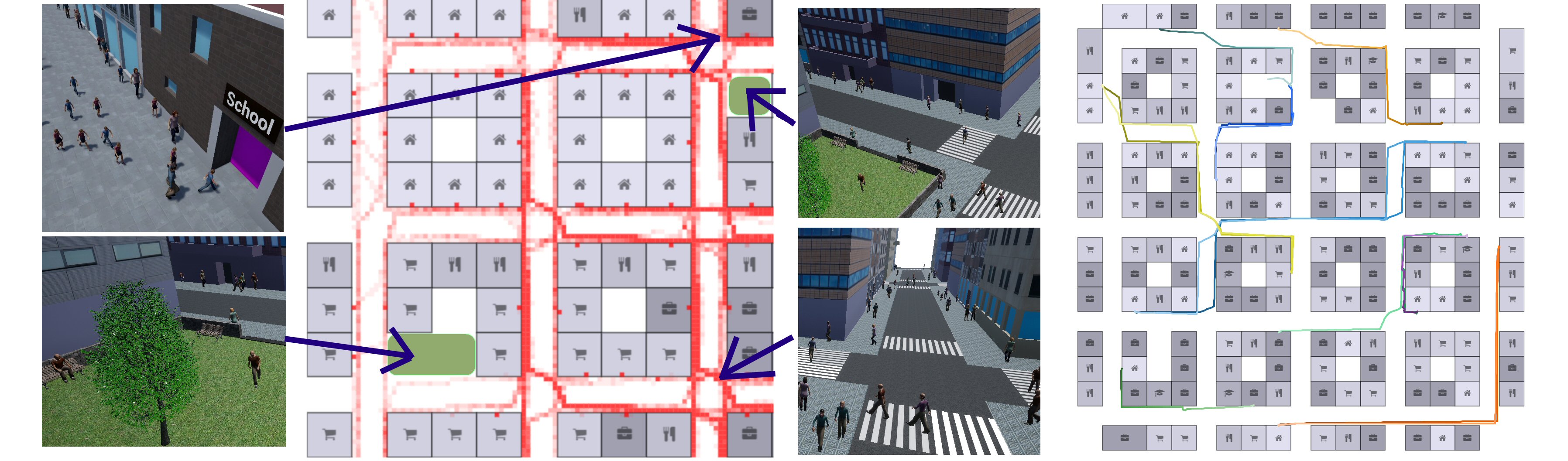}
		\caption{Example of a city with a generated population, highlighting different behaviors such as going to work, bringing children to school, or sitting in a park. The left heat-map shows the population whereabouts through a day, and the map on the right shows some consistent example trajectories. Individual agendas are fully created in a procedural manner, exhibiting consistent behavior, e.g., an adult leaves his/her house to take his/her children to school and then go to work. In the afternoon that same adult will go back to the previous school, pick up the same two kids and go back to the initial house.}
		\label{fig:Teaser}
		%}
	\end{teaserfigure}

	\begin{abstract}
		Authoring realistic behaviors to populate a large virtual city can be a cumbersome, time-consuming and error-prone task. Believable crowds require the effort of storytellers and programming experts working together for long periods of time. In this work, we present a new framework to allow users to generate populated environments in an easier and faster way, by relying on the use of procedural techniques. Our framework consists of the procedural generation of semantically-augmented virtual cities to drive the procedural generation and simulation of crowds. The main novelty lies in the generation of agendas for each individual inhabitant (alone or as part of a family) by using a rule-based grammar that combines city semantics with the autonomous persons' characteristics. Real-world data can be used to accommodate the generation of a virtual population, thus enabling the recreation of more realistic scenarios. Users can author a new population or city by editing rule files with the flexibility of re-using, combining or extending the rules of previous populations. The results show how logical and consistent behaviors can be easily generated for a large crowd providing a good starting point to bring virtual cities to life.
	\end{abstract}

	
	
	
	\maketitle
	
	%---------------------------------------------------------------------
	%---------------------------------------------------------------------
	%---------------------------------------------------------------------
	\section{Introduction}
	
	%NURIA: VOY QUITANDO DEL PDF LOS COMENTARIOS QUE YA TENEMOS EN CUENTA O PARA VER REALMENTE LO QUE OCUPA EL PAPER, ASÍ QUE MIRAR LOS COMENTARIOS NUESTROS AQUI Y NO EN EL PDF ;)
	
	%\Otger{malgrat que el nostre objectiu es videojocs/cine, enfatitzar us per a urban planning (e.g. carrers peatonals d'una direccio a Madrid (anonimitzat!))}
	
	%\Otger{results: comparacio amb random (heatmap), flow 1 familia, comparacio amb e.g. crowd patches em tema facilitat d'authoring. Estimar temps de crear un arxiu de regles / fusionar-ne dos vs. creació amb crowd patches (o altres sistemes, Nuria?). Estimar per un exemple concret de ciutat? Destacar flexibilitat: exemples concrets, e.g. facil forçar la gent a anar a tal hora el diumenge a l'ajuntament a votar, facil de canviar; o simulació d'evacuació (e.g. amenaça catàstrofe inminent). Destacar comportaments amb grups d'agents -dinamics- (altres sistemes, o be cada agent va sol, o grups fixes per tota la simulacio). Destacar objectiu: que tot tingui sentit, facilitat de obtenir comportaments rics i/o emergents.}
	
	%\Otger{Com a treball futur (que es pot comentar @ discussió), dir que podem fer que la gent tingui estats emocionals (e.g., he deixat els nens al cole però arribo tard a la feina), que pot permetre estudiar probablilitats d'accidents, embussos, problemes, etc. A més podriem afegir límts de velocitat (tràfic), etc. Possibilitat aprenentatge automatic (si un dia faig tard, l'endema surt amb mes temps de casa)}
	
	% ======= What is known ======
	Large virtual environments can be easily found on the Internet, and can be visualized with inexpensive VR setups. Having a believable crowd requires the work of a dedicated team of expert artists, story writers and animators working for months to achieve the desired virtual population. There are currently several tools to ease the generation of virtual humans, such as Mixamo~\cite{Mixamo}, MakeHuman~\cite{MakeHuman} or Autodesk Character Generator~\cite{AutodeskCharacterGenerator}. These tools provide rigged and textured skeletal meshes along with a variety of animation clips. Game engines also include local movement algorithms and path finding, but the task of assigning goals to the characters is left to the user, and those tasks are typically limited to locations without any further meaning.
	
	% ======= What is unknown ======
	The one thing missing in most virtual environments are believable, purposeful virtual humans. The main reason is that there is still a significant lack of tools to populate such environments in an easy, meaningful and semi-automatic manner. Essentially, the process of getting those virtual humans fully integrated in an environment, being simulated and showing a purpose as they wander the virtual worlds, is still a challenge. 
	
	% ======= our burning question + contributions ======
	In this paper, we propose a new procedural framework to rapidly generate large semantically augmented cities, to then drive the behavior of a generated crowd of virtual citizens. This bridges the existing gap between the individuals' behaviors and the city they populate. Our system for Procedural Crowd Generation (PCG) allows users to create behaviors for large crowds by writing rule sets containing only a handful of rules, thus providing high scalability, while achieving diversity, flexibility, and re-usability.
	
	The main novelty of our approach is the procedural generation of dynamic agendas to simulate the high level behavior of the citizens of a city. There has been a large amount of work on procedural generation of cities, and also in the crowd simulation field. However, there is a significant lack of integration between those two research areas, and typical individuals in a simulated crowd lack higher level agendas like the ones we are proposing here. Our method allows to generate behaviors that are consistent, meaning that, if we follow any character though a day, we will observe a logical sequence of actions, e.g., if an adult leaves home with two kids, it should leave them at school before going to work, and it should pick up the same two kids in the afternoon and go back to the same house where they left from in the morning. This can greatly enhance the believability of secondary characters in video games or story telling, but most importantly it will generate consistent trajectories and schedules that could be useful for urban planing.
	
	Our semantically enhanced cities can be modified by the user on the fly (e.g., change location of objects or buildings), or by altering the rule files (e.g., modifying size, type or distribution of buildings) and without requiring any additional work by the user, the system will automatically adjust agendas and behaviors. Similarly, individual's agendas are generated with dynamic procedural rules, meaning that we allow rules to be triggered during simulation time instead of being fixed by the generation of the crowd. Note how our work does not compete against current crowd simulation models (those dealing with steering, planning or following formations), since it represents a higher level of abstraction and, thus, it could be combined with any current model to simulate path finding and local movement.
	
	%With this, we are aiming at procedurally generating higher level interactive and dynamic behaviors, 
	%with little effort from the user. The procedural generation of cities with embedded semantics is a necessity of our system to run such agenda generation.
	
	The main contributions of this work are: (1) the procedural generation of semantically-augmented virtual cities; % with embedded information regarding land usage, zones of interest, and interactable objects; 
	(2) %an easy-to-use 
	a crowd authoring tool based on a \emph{rule-based grammar} with dynamic agendas; % and custom or real-world population data; (3) the procedural generation of dynamic agendas for each individual with their daily schedules; 
	and (3) a procedural crowd simulation method. 
	%NURIA: NO DIGAMOS EASY TO USE DEL AUTHORING TOOL ASÍ NOS EVITAMOS QUE NOS PIDAN HACER USER STUDIES PARA PROBARLO, DIREMOS QUE SE USAN RULE FILES QUE SE PUEDEN EDITAR PARA RAPIDAMENTE GENERAR OTRA POBLACION.
	
	\change{Note that the scope of this research does not involve aspects such as character animation, path finding, group movement, obstacle avoidance, or collision resolution. In our system, those features are handled by the default functionalities of the game engine used for the implementation.}
	
	%NURIA ESTO LO QUITO AHORA Y LO DEJO PAR RESULTS, SON IMPLEMENTATION DETAILS NO IMPORTANT PARA SIGGRAPH:
	%Our framework has been integrated into Unreal Engine 4~\cite{UnrealEngine4}, but designed with enough flexibility to be easily adaptable to be used within any other engine.
	
	%---------------------------------------------------------------------
	%---------------------------------------------------------------------
	%---------------------------------------------------------------------
	\section{Related Work}
	%---------------------------------------------------------------------
	%---------------------------------------------------------------------
	\subsection{Procedural urban modeling}
	Procedural modeling has been used for both urban and non-urban environments, generating very rich geometry requiring only a very small effort and time from the user~\cite{Parish01,Wonka2003,muller2006procedural}. One of the most prominent approaches for the procedural generation of cities consists on the creation of a 2D map of a city (with building lots, road networks, etc.), followed by a procedural generation of buildings on the lots described in it. 
	On the generation of the exterior geometry of buildings, a powerful approach and probably the most popular one is Computer Generated Architecture (CGA)~\cite{muller2006procedural}. It defines a rule system based on parameterized grammars (\textit{shape grammars}). It allows writing rules on how to generate the buildings, refining the details through grammar productions, and allowing for parameters and the use of pseudo-random values to generate variations. This method has been subsequently improved (CGA++~\cite{schwarz2015practical}), and it is being used in the commercial software CityEngine~\cite{CityEngine}. The interested reader is referred to the surveys by Watson et al.~\shortcite{Watson08}, and Vanegas et al.~\shortcite{Vanegas09} for an in-depth review of the state of the art literature in urban modeling. Recently, Benes et al.~\shortcite{Benes:2014:PMU:3071829.3071839} presented a procedural method based on the key idea that a city cannot be meaningfully simulated without taking its neighborhood into account, using a traffic simulation plus information about a city neighborhood to grow both major and minor roads. 
	
	Statistical information can be gathered to allow the user to create outdoor terrains by learning the distribution of trees, grass, rocks, etc. constrained by the terrain slope~\cite{emilien2015worldbrush}. The user can then paint and populate new terrains copying the object distribution. Hendrikx et al.~\shortcite{SurveyProceduralGames} discuss in a survey several procedural generation methods for different kinds of content or aspects.%, and provide some examples in the field of video games. The discussed content types include generation of urban environments and buildings.
	
	%---------------------------------------------------------------------
	%---------------------------------------------------------------------
	\subsection{Inverse urban problems}
	Garcia-Dorado et al.~\shortcite{Garcia-Dorado10.1111:cgf.12329} presented a technique to enable a designer to specify a vehicular traffic behavior, in such a way that their system is able to inversely compute what realistic 3D urban model yields that specific behavior. Similarly, Feng et al.~\shortcite{feng2016crowdlayouts} presented a system to design mid-scale urban layouts by optimizing the parametrized model with respect to some metrics related to crowd flow properties: mobility, accessibility, and coziness.
	Peng et al.~\shortcite{peng2016funcNetwork} presented a system that generates networks for design scenarios like mid-scale urban street layouts and floor-plans for office spaces, where the user specifies an input mesh as the problem domain along with high-level specifications of the functions of the generated network, such as preference for interior-to-boundary traffic, interior-to-interior traffic, networks with specified destinations (i.e., sinks) on the boundary and local feature control, such as reducing T-junctions or forbidding dead-end.

	\subsection{Simulation-based approaches}
	Most of the current work on using simulations to control the agent behaviors aims to provide quick and easy ways for the user to specify trajectories, destination points or densities. 
	Crowdbrush was proposed as an interactive tool to easily edit crowd behavior showing basic animations~\cite{Ulicny04}. Yersin et al.~\shortcite{Yersin:2009:CPP:1507149.1507184} presented Crowd Patches, which provide high performance by pre-computing paths. Agents are moved as if they were trains on tracks. Therefore, if we observe an agent or a small part of the environment for a certain time, we will soon notice repetitions. Extensions of crowd patches permit authoring densities, and flows, editing crowd patches and even alternating patches to mitigate repetitions\cite{jordao2014crowd,jordao2015crowd}. However, there is still no link between the environment and the individual's behavior, other than points where agents appear or disappear. 
	
	Kim et al.~\shortcite{kim2014interactive} presented a similar system using Cage-based deformations for the interactive manipulation and animation of large-scale crowds.
	Navigation Fields \cite{patil2011directing} allow the user to sketch directions to deviate the movement of the agents from their original path towards theirs goals. However, there is no higher level planning or intentions. 
	
	%The above systems deal with the overall movement of individuals, but treat them as empty boxes without higher level planning or intentions. 
	
	Badler et al.~\shortcite{Badler2000parameterized} proposed a parameterization of agent actions that consists of rules parametrized by the participants (objects and agents), and was employed to translate high-level orders from natural language into low-level tasks. Allbeck and Badler~\shortcite{allbeck2002toward} presented one of the first systems to take into account character believability, personality, and affect, by using a parameterized action representation, which allows an agent to act, plan, and reason about its actions or other agents actions.
	Li and Allbeck~\shortcite{LiA11} proposed a system to provide purpose to a given population by using an agent-based simulation to create virtual populations endowed with social roles. %, which help establish behaviors of each agent. 
	Semantic information can be embedded in objects for the purpose of interaction between virtual humans and objects \cite{kallmann1999modeling}. The work by Pelkey and Allbeck \shortcite{pelkey2014populating} focuses on semi-automatically assigning semantic affordances to objects. However, most semantic tagging of objects still requires some degree intervention of the user and tagging by hand. 
	Recently, Parameterized Behavioral Trees have been extended to handle small groups of agents for the purpose of storytelling. The user can drag\&drop actors and actions to create narratives (M. Kapadia~\shortcite{Kapadia:2016:MIG}). 
	The same year, Krontiris et al.~\shortcite{Krontiris16} presented an activity-centric framework for authoring heterogeneous virtual crowds in semantically enriched environments, where the locations are labeled as environmental attractors and agent desires are used to compute influence maps, which ultimately drive their behavior.
	The difference with our work is that the story or role for each individual still needs to be created manually and does not emerge in a procedural way from a simple set of rule files. %Also, our work offers greater flexibility and dynamic agendas, as it does not require defining a role for each slightly different behavior.
	
	Jorgensen et al.~\shortcite{jorgensen2014space,jorgensen2015scheduling} presented scheduling algorithms to plan the activities of agents, but their work was limited to a per-agent specification of such activities. Semantic-driven crowds can also be created with ontology-based models~\cite{Pavia2005}, at the cost of requiring to manually define multiple profiles.
	Kraayenbrink et al. \shortcite{kraayenbrink2012semantic} proposed a system based on satisfying agent ambitions and desires, using heuristic formulas and optimization, which limits their method to small crowds. 
	Bulbul et al.~\shortcite{bulbul2010populated} proposed a system that retrieves geolocalized data from social media sources like Twitter and Instagram to populate virtual cities. Recently, Kapadia et al.~\shortcite{Kapadia:2016:SCA} presented a system called CANVAS, a computer assisted visual authoring tool for synthesizing multi-character animations from sparsely-specified narrative events. In general, we can say our work presents a more powerful and flexible way of including semantics and defining complex relationships, allowing our environments to scale up, and easing the effort of creating diversity of behaviors for large crowds without much user intervention.
	
	We refer the reader to the book by Kapadia et al.~\shortcite{kapadia2015virtual} for more information about methods, simulation and control of virtual crowds. 
	
	%The idea of using procedural techniques to generate crowds is not new, however, none of these works includes procedural simulation, which we consider being of crucial importance to achieve a realistic behavior of a city and its citizens.
	The work by Maim et al.~\shortcite{VAST:VAST07:109-116} uses a procedural method that combines city generation (using CityEngine) with semantics to trigger basic behaviors with their corresponding animation, such as "look at", "walk into", or "slow down". However, there are no higher level agendas or planning involved in such simulation. %Our technique goes beyond basic behaviors and into higher level agendas with an emphasis on reusability and extensibility.
	Katoshevski et al.~\shortcite{JTLU333,KatoshevskiCavari2011131} and Rasouli and Timmermans~\shortcite{Rasouli201479} developed a set of tools, based on the previous work by Arentze and Timmermans~\shortcite{Arentze_1albatross}, to analyze the impact of different urban shapes on activity-travel patterns, together with the evolution of land use, in a context related to pollution emission. For this, they use a set of heuristics to transform statistical socio-economic data into decision trees to drive the agents behaviors. On the other hand, Torrens' work~\cite{Torrens07,TGIS:TGIS1261,Zou12,Torrens20121}, %although very interesting on its own, 
	is mainly based on using GIS data to identify possible targets for the agents, but there is no connection with the city beyond these simple identifications. %, and the agents do not have an agenda or similar. 
	Likewise, the MATSim~\cite{MatSim} project uses census information (only available for real cities) or random location selection for the city integration. In general, all these systems randomly select locations and assign simple, fixed plans for the agents (e.g., Home-Work-Home) based on probability distributions. They cannot, thus, reach the flexibility and control presented by our system.
	
	%---------------------------------------------------------------------
	%---------------------------------------------------------------------
	\subsection{Simulation in commercial software}
	Games try also to show interesting crowds mostly through a large variety of animations and appearances, however high level behaviors are typically scripted or driven by finite state machine that are created by the programmer for each NPC or at groups of NPCs. Assassin's Creed: Brotherhood~\cite{AssassinsCreedBroderhood} implemented a simple day/night variation on the character animations and appearance. Other games like Hitman: Absolution~\cite{Hitman} use a finite state machine where states are changed by environmental or player-generated situations, but the overall actions are manually scripted. A more sophisticated crowd behavior can be observed in Assassin's Creed Unity~\cite{AssassinsCreedUnity1,AssassinsCreedUnity2}, where the characters are uniquely shepherd, following randomly generated closed paths, in a way such that the overall behavior is fully deterministic. A more recent example comes from the Watch Dogs 2 game~\cite{WatchDogs2}, which simulates emergent behavior by integrating manually placed, scripted attractors, with a fuzzy-logic-based reaction system that encodes the possible responses of NPCs to events in the game. These solutions work well at first, but players eventually notice the repetitiveness of behaviors, the lack of long term goals, or the overall inconsistencies (e.g. characters that disappear in a location and magically appear in a different one).
	Final Fantasy XV~\cite{Imamura:2016:FFX:2897839.2927449}, as many other modern games, uses scripted behaviors for cinematic sequences, and relies on established techniques like behavior trees for in-game situations, which is an interesting solution that, unfortunately, requires the designers to predict and code any possible behavior an agent may require. Also, again players are able to observe repetitions and patterns, resulting from the limitations of using a single behavior tree (with many branches and some randomization) for an entire crowd.
	Creation Engine \cite{CreationKit}, a tool used in games such as Skyrim and Fallout 4, is a system in which NPCs can be scheduled to perform actions at certain times of the day (e.g., go to a place, eat, sleep, etc.), dynamically filling details such as finding the nearest available food item. While this results in rich and consistent individual behaviors that adapt to the world, it requires a specification of the agent tasks by hand, which makes it non viable for large crowds.
	
	Finally, it is important to mention that many commercial applications like  Maya~\cite{Maya} have specifically tailored plug-ins for procedural crowd simulation, like Miarmy~\cite{Maya-Miarmy} and Golaem~\cite{Maya-GOLAEM}, which provide some degree of control over the generated agents but with an aim at short scenes.
	
	In spite of all those efforts, there is little correlation between the agent behaviors and the city itself, except for representing physical barriers that the agents cannot go through. In this work we propose to bridge this gap, establishing a direct and clear relationship between the agents and the city they live in. Moreover we tackle the challenge of generating agents with dynamic agendas and higher level plans, to achieve a simulation that goes beyond agents wondering aimlessly around the environment.
	
	%-------------------------------------------------------------------------
	%-------------------------------------------------------------------------
	%-------------------------------------------------------------------------
	\section{Overview: Procedural Crowd Generation (PCG)}
	
	\begin{figure*}[h!]
		\centering
		\includegraphics{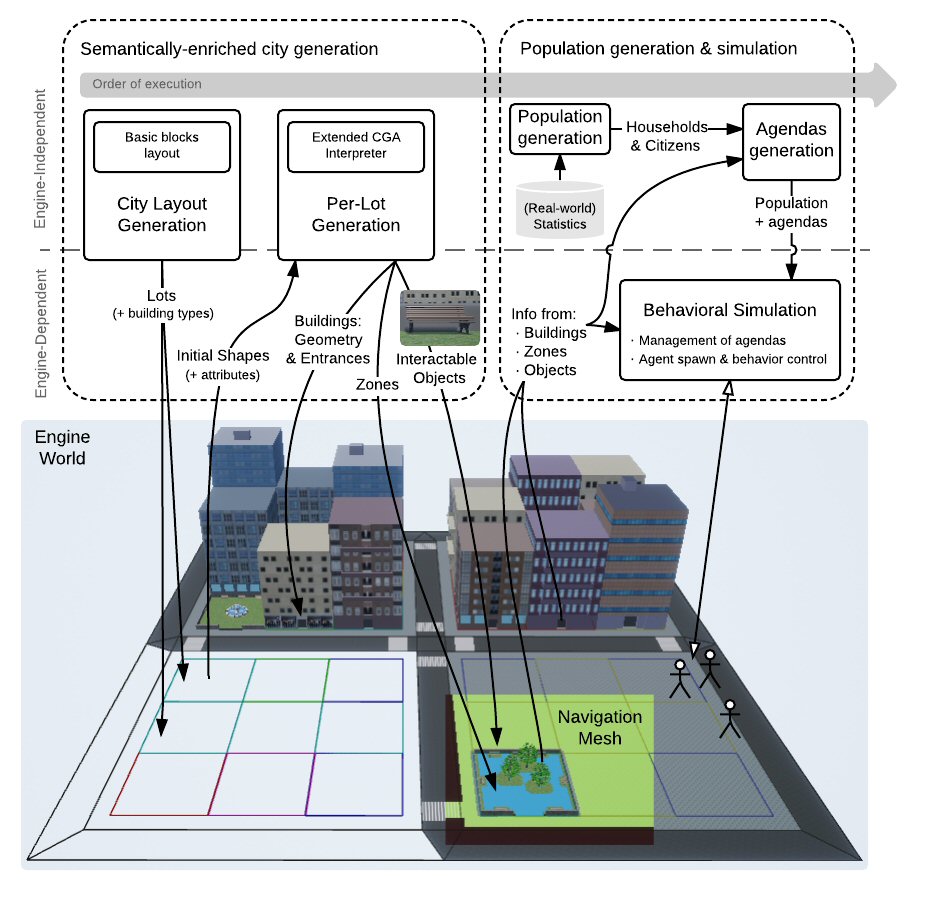}
		\caption{Overview of the framework for generating populated virtual cities. Arrows represent the input and output data of each module.}
		\label{fig:SystemOverview}
	\end{figure*}
	
	Our procedural framework consists of two main modules: the city and the population modules. %plus some sub-modules to keep dependencies low and offer reusability of code. 
	The former deals with the generation of the semantically-enriched cities, and the latter with the generation of the population and its behavior.
	
	%NURIA: ESTOY QUITANDO BLA BLA BLA QUE LUEGO SE REPITE EN LAS SUBSECCIONES
	%The first module contains the generation of the city, by first
	The city generator starts by creating its layout (i.e., the set and shape of the lots), and then the buildings for each lot. %For the former we have used a simple city block generator. 
	%
	%For the generation of buildings, a custom implementation of CGA (Computer Generated Architecture~\cite{muller2006procedural}) is used, which generates building geometry by executing a set of grammar-based rules whilst augmenting the results with semantic data.
	%He modificado el parrafo anterior por este nuevo, porque creo que asi destaca mas que es lo que hemos creado nuevo:
	The generation of building geometry and layout is based on CGA (Computer Generated Architecture~\cite{muller2006procedural}). However, we have implemented a specifically tailored version of CGA, respecting the original CGA syntax, but extending the basic grammar-based rules in order to include semantic data that augments the information associated with each element in the city.

	The population generator, is divided in three stages: the generation of the city population, the generation of each individual \emph{agenda}, and finally the behavioral simulation. The population is defined as a set of households ("families") with members of various ages and genders. The \emph{agendas} are generated using a system of rules loosely inspired on CGA grammars, extended with a novel set of functions, variables and operations that have been specifically designed to handle dynamic events and flexible agendas.
	%with the goal of providing flexible \emph{agendas} and the ability to handle dynamic events.
	
	The city and crowd modules are designed separately, and use an engine world (or game world) as the main structure for exchanging information, by creating entities or components of different kinds as results, or querying and using them as input. With this design we achieve greater flexibility, by allowing the user to perform modifications by hand if needed after any of the generations, or experimenting in each step with different parameters or input files to immediately evaluate its impact in the city dynamics.
	%until the desired results are achieved. %NURIA he quitado esto porque da la sensación de que el usuario simplemente puede modificar eternamente hasta que tiene la suerte de que le guste el resultado, sin una clara relacion entre lo que toca y lo que acaba sucediendo, creo que decir que nos permite evaluar inmediatamente el impacto de los cambios, hace el metodo mas fuerte.
	Figure \ref{fig:SystemOverview} shows an overview of the different modules of the system. %, their dependencies, and how they interact with each other.

	%-------------------------------------------------------------------------
	%-------------------------------------------------------------------------
	%-------------------------------------------------------------------------
	\section{City generation}
	
	The generated cities consist of both the geometry and the \emph{semantic} information that is used later on to drive the citizens' behavior. As opposed to previous work where cities and population are two independent elements with no connection other than a navigation mesh (\emph{navmesh}) with manually annotated information,  %to steer the crowds, 
	our framework provides a direct link between these two entities. This drastically reduces the effort of embedding behaviors into the citizens. %In order to achieve our goal, the procedurally generated cities are automatically augmented with semantic information. 
	
	Cities are generated in two steps: (1) the layout is generated, consisting of the building lots, roads, etc.; and (2) the geometry for each element is generated, in a procedural way. % using rule-based modeling.
	
	The city layout generator creates a set of rectangular city blocks with square-like lots. %The dimensions can be tuned by the user with a set of input parameters such as: 
	The user can tune the city by editing the number of blocks, number of lots per block, block shape and the aspect ratio. This could be extended with more sophisticated strategies, as the creation of the urban environment is completely independent of the other parts of our framework. However, a more sophisticated layout lies outside the scope of this work.
	
	The generation of the building geometry is based on Computer Generated Architecture (CGA)~\cite{muller2006procedural}, more specifically on the syntax version used in CityEngine~\cite{CityEngine}. % and specifically the variant used in the commercial CityEngine system~\cite{CityEngine}. 
	%NURIA HE QUITADO ESTO PORQUE ME PARECE DEMASIADO RELACIONADO CON CITY GENERATION Y NO ES EL FOCUS DE ESTE TRABAJO.
	%This brings us all the power of this approach, and even though the current examples use simple buildings with low detail, artists can unlock its full potential to generate highly-detailed geometric models with relatively small work.
	The basic syntax of a \emph{CGA rule set} consists of a plain text file containing a list of rules. Each rule is on the form of the rule name and an optional list of arguments, followed by an arrow, and then the successor. The successor is a list of one or more rule names and operations. These items will be executed sequentially, and may also be called with arguments. An example of a rule is:
	
	\texttt{{ A --> B  C(arg1, arg2, ...)  op(arg) }}
	
	Some of the operations require the presence of a selector block, which specifies a different successor for various cases or conditions. The decision of whether to execute each item depends on how the specific operation interprets the selector. For this case the syntax is:
	
	\texttt{op(args) \{ selector1: successor1 | ... \}}
	
	For a more complete reference of the language syntax and control structures, we refer the reader to the CGA reference manual provided by CityEngine~\cite{CityEngine}. Although we use a custom implementation of the CGA interpreter, we decided to follow the same syntax in order to support most existing rule files and make them easily adaptable to our system. Listing \ref{code:BuildingRuleFileExample} shows an example of a rule set with semantics. 
	%NURIA: QUITO ESTO, DEMASIADO DETALLE DE IMPLEMENTACIÓN A ESTA ALTURA DEL PAPER
	%Note that redundant white spaces and line breaks are ignored (except inside strings), and expressions, operands and the call syntax are the same as in the C/C++ programming languages, with no semicolons.

	Our contribution lies on extending the CGA specification %for the generation of the geometry, 
	to embed different semantic elements during the geometry generation. A key element introduced in our work is the possibility of defining within the lots, a single or multiple entrances, which can be used as entry/exit locations to plan individual's trajectories. This provides the flexibility to define, for example, a store on the ground level and an entrance to the residences in the upper floors.
	
	Our system can also specify zones of interest inside the lots, such as park zones. This allows for the flexibility of having, for example, a small building with a little interior park within the same lot, as opposed to having the entire lot occupied exclusively by a single large building or by a park.
	
	CGA can be visualized as generating a tree of oriented bounding boxes which are not restricted to the parent box and have attached properties (e.g. material), and where the tree leaves correspond to the actual generated geometry. Taking advantage of this recursive nature, we allow subsets of the generated geometry (i.e. subtrees) to be tagged with user-defined tags. This can be used for example to mark some of the generated geometry as a bench or a lamppost.
	
	%Lots are assigned types, which are currently: house, shop, workplace, school, leisure, park
	%Each lot can have several zones and entrances
	%Lots-> pueden tener una o mas zone y entrances. En cada lot podemos meter (building types):
	%lots types: house, shop, workplace, school, leisure, park
	%building types: todo lo anterior menos park.
	
	%Therefore, all the information regarding city structure and organization is created procedurally following the input files, where the user can configure the characteristics of the city. 
	
	Formally, our extensions to CGA include the introduction of two new operations which affect the current \textit{CGA scope} (i.e. the current ``bounding box''): (1) the \texttt{entrance} operation creates an entrance to a building of the user-defined type, at the origin of the current scope; and (2) the \texttt{zone} operation defines a zone of interest of a user-defined type spanning the current scope. Additionally, we include a new annotation \texttt{@object} allows tagging a generated CGA subtree as a separated object (specification details can be found in the Appendix).
	
	Listing \ref{code:BuildingRuleFileExample} shows a simple but complete example of a CGA rule file extended with all these semantic elements. It splits the lot in two along the X axis, generating a building with an entrance on one side, and a zone marked as a park, containing some benches on the opposite side.
	
	The most important aspect of our city generator is that, as the city is being automatically created, it tags different elements and zones with semantics (e.g. buildings, zones of interest or interactable objects) that will be evaluated by the crowd generation and simulation module to drive the city dynamics. This city semantics are also combined with the final navigation mesh %, thus providing a navigation mesh that has been automatically enhanced with relevant information, which will 
	to drive path finding and behaviors (i.e destination positions, find the closest park to rest, or find an empty bench in the park).
	
	%\NURIA: HE AÑADIDO UNA LINEA ARRIBA PARA PODER CARGARME TODO ESTE PARRAFO QUE RESULTABA REPETITIVO.
	%Once the semantically augmented city is created, it is processed by the simulation engine (or game engine), which will store the generated semantic components (e.g. buildings, zones of interest or interactable objects) to be used by the simulation, and to generate the navigation mesh taking into consideration that information. 
	
	To further enhance the flexibility of our system, %there is an additional layer of interaction before creating the population. This layer allows 
	%we allow 
	the user can inspect and edit the generated city by tuning the parameters and the rules, before creating the population. % until the desired result is achieved.
	At this point, it is possible to further refine the city semantics or the content itself %, by manually editing the results 
	(e.g. to add specific unique landmarks such as monuments, or copy/move/paste objects). %This provides a fine level of control over the final result.
	This combines the advantages of fully automatic generation with the flexibility of fine-tuning when needed.
	
	%\newpage
	
	\begin{lstlisting}[frame=single,style=CGA,caption=Example of a CGA rule file augmented with semantics,label=code:BuildingRuleFileExample, basicstyle=\small] %,float=*]
	
	@StartRule
	Lot    -> split(x) { 8: Shop | ~1: Park }
	
	Park   -> zone("park")
	split(x) { ~1: NIL | 0.4: Benches }
	Benches -> split(z) {~2: Bench | ~2: NIL }*
	
	@Object("bench")
	Bench  -> extrude(0.5)
	Shop   -> extrude(5)
	comp(f) { front: Facade | all: Wall }
	Facade -> split(x) {~1:Wall | 1.5:DoorV | ~1:Wall}
	DoorV  -> split(y) { 2: Door | ~1: Wall }
	Door   -> [ t('0.5, 0, 0) entrance("shop") ]
	extrude(-0.2)
	color(0.2, 0.2, 0.2)
	comp(f) { back: NIL | all: Wall }
	
	\end{lstlisting}

	%--------------------------------------------------------------------
	%--------------------------------------------------------------------
	%--------------------------------------------------------------------
	\section{Generation of the population}
	
	A real society is not formed by a collection of independent individuals. Societies are formed by families and friends, and most of our daily routines evolve around our relationships with a number of individuals that belong to the same group. Therefore, at the core of our population generation module we have the \textit{household}, $H$ (which typically includes all family members living in the same house). Households can consist of one or more individuals, $\iota$, so we define a family as a group of individuals: $H=\sum^{n}_{i=1}\iota_i$, and our overall population, $\Pi$, will emerge from a collection of households $\Pi=\sum^{m}_{j=1}H_j$. Just as in the real world, our collective population whereabouts will emerge from the combination of the individuals' needs with the family schedules.
	
	\subsection{Population}
	
	To generate realistic households, we use real-world data, extracted from a city hall open-data service (\cite{Annonymous}). This data contains pairs of a denotative descriptor string of the household type (e.g. ``one adult woman with two minor children'') and the total number of occurrences of this pattern in the real city. For each pair, we derived a household pattern by mapping each semantic descriptor into a set of details, such as: number of adults, sex or number of kids. The numbers of occurrences are then normalized and used as the probability of choosing each pattern during the procedural generation of the virtual population, by applying a cumulative density function during the sampling process. The virtual population is distributed across the city based on the capacity of each building (i.e. number and size of apartments in a building).

	%--------------------------------------------------------------------
	%--------------------------------------------------------------------
	\subsection{Agendas}
	
	The agendas represent the sequence of tasks and goals that individuals will perform during their day. Just like in real life, individuals' agendas depend strongly on the composition of their households. For example, adults with kids may need to take the kids to school before going to work.
	%Just like in real life, agendas have tasks with different levels of priorities, different relevance, some with strict times and other that can be swift around our day or even moved to a different day. Our virtual agendas need to capture all the richness and variability of the real world. 
	The advantage of using agendas to drive the citizens' behaviors is that it eases the authoring of the virtual population by allowing the user to define high-level tasks under an intuitive model.
	
	In our system we propose a rule-based modeling system inspired by the CGA, but oriented at the procedural generation of dynamic agendas. This allows us to define how to generate automatically all the agendas as opposed to authoring each one individually, while allowing for some variation based on pseudo-randomness. 
	Our agenda generator is thus scalable, heterogeneous and adjustable.
	
	%--------------------------------------------------------------------
	\subsubsection{Generation of the agendas}
	
	The agenda generation is performed by executing a \emph{Rule Set}, similarly to a CGA rule set in structure and syntax, but extended to support a different set of built-in global variables, functions and operations. In addition to this, for the generation of agendas there is no rule context (i.e. no analogue to CGA's \textit{shape} and its attributes). The complete details of the syntax, functions and operations can be found in the Appendix. In our system, functions typically allow the user to query the city environment (e.g. distances, find elements, etc.)% and have no side effects
	, whereas operations do not return values but perform some effect, such as creating tasks in the agenda.
	
	The generation of the agendas starts at the household level, instead of for each individual. This allows us to have access to information of the household and interleave entries in the individual's generated agendas as required. Thus, we introduce the concept of \textit{focusing} a person of the household, which limits whose agenda is being affected. During the generation of the agendas for a household, a single member at a time can be \textit{focused}, so that all agenda-modification operations affect exclusively its agenda, having no impact on the agendas of other members of the same household.
	
	A novel and powerful feature of the operations and functions in \emph{PCG}, is having parameters that can be a \textit{predicate} instead of a value. A \textit{predicate} is an expression that is not evaluated before the function is called, but evaluated inside the function using a different context (e.g. changing the focused members). The main application of this is to evaluate conditional expressions (e.g. \texttt{age > 18}) for every household member to make decisions during the generation.
	
	%In addition, we use numeric IDs to reference agents, households or buildings, and use them as values received from or passed as argument to functions or operations. 
	
	As the system is based on a parameterized grammar, rule calls can have parameters assigned to them, which increases the reusability of the user-written rules. Moreover, the \emph{Rule set Interpreter} can process imports of multiple \emph{rule sets} for the generation of a population. This allows our framework to provide both flexibility and reusability similarly to what CGA offers for the generation of the geometry: the user can create new populations from combinations of previous ones, or easily extend a previous model adding just a handful of new rules. 
	
	A key difference with CGA is that our procedural agenda generation does not handle the agenda by dividing it in portions in a recursive way (i.e. recursive refinement), but as rules are triggered they can introduce entries at any time of the day. Therefore, adding new agenda entries may result in conflicts, and to remove the burden on the user when creating the population, our system handles these conflicts automatically. In case of a collision between agenda entries, the newest one %(in execution order) 
	takes precedence: the overlapping entries in the agenda are either split in two, trimmed, or deleted, as needed to fit the newer task in the agenda. The user is thus only responsible for ordering the calls to rules to establish these priorities.
	
	%As priorities are assigned based on the order of the execution, the only thing that the user needs to take care of is to indicate first those tasks with smaller priorities. %(AS FUTURE WORK WE WILL ALLOW A SYSTEM TO INTRODUCE TASKS PRIORITIES)\\
	
	Algorithm \ref{alg:AgendaGeneration} formalizes the process used to generate the agendas. Notice how the evaluation context is kept local to the current rule, i.e. modifying it will affect the remaining of the rule as well as any further rule called from it, but will not affect the context of the calling (parent) rule. The evaluation context chiefly contains: the current household, the currently focused person -if any- and information about it, the current mask of days to modify, and a set of local variables (including the arguments passed to the current rule call). Note that the generated agendas themselves are not part of this context, and any modifications to them will be seen by any posterior rule, including any parent one, effectively working as a ``global'' variable.

	\begin{algorithm}
		\SetKwFunction{FClone}{Clone}
		\SetKwFunction{FEvaluateArgs}{EvaluateArgs}
		\SetKwFunction{FInsertAgendaTask}{InsertAgendaTask}
		\SetKwFunction{FApplyOperation}{ApplyOperation}
		\SetKwFunction{FApplyRule}{ApplyRule}
		\SetKwFunction{FGenerateAllAgendas}{GenerateAllAgendas}
		\SetKwFunction{FDelayedExecutionTask}{DelayedExecutionTask}
		
		\SetAlgoLined
		\DontPrintSemicolon
		
		\SetKwProg{Fn}{Function}{:}{}
		\SetKw{kwNew}{new}
		
		\Fn{\FGenerateAllAgendas}{
			$r$ $\leftarrow$ rule marked with $@StartRule$ \;
			\ForEach{$h$ $\in$ households}{
				$ctx$ $\leftarrow$ \kwNew GenerationContext \;
				$ctx \ldotp household$ $\leftarrow$ $h$ \;
				\FApplyRule($ctx$, $r$) \;
			}
		}
		
		\;
		
		\Fn{\FApplyRule{$ctx$, $rule$}}{
			
			\ForEach{$s$ $\in$ $rule \ldotp successor$}{
				arglist $\leftarrow$ \FEvaluateArgs{$s \ldotp args$} \;
				\uIf{$s$ is a rule call}{
					$ctx'$ $\leftarrow$ \FClone{$ctx$} \;
					$ctx' \ldotp variables$ $\leftarrow$ $ctx' \ldotp variables$ $\cup$ arglist \;
					\FApplyRule{$ctx'$, $s \ldotp rule$} \;
				}
				\uElseIf{$s$ is an operation call}{
					$ctx$ $\leftarrow$ \FApplyOperation{$ctx$, $s \ldotp op$, arglist} \;
				}
			}
		}
		
		\;
		
		\Fn{\FApplyOperation{$ctx$, $op$, $arglist$}}{
			\eIf{$op$ = "DelayedRule"}{
				$t_0$ $\leftarrow$ arglist[0] \;
				$t_1$ $\leftarrow$ arglist[1] \;
				$rule\_name$ $\leftarrow$ arglist[2] \;
				\ForEach{$p$ $\in$ $c \ldotp household \ldotp members$}{
					\uIf{$p$ is focused}{
						q $\leftarrow$ \kwNew \FDelayedExecutionTask{$rule\_name$} \;
						\FInsertAgendaTask{$p \ldotp agenda$, $ctx$, $t_0$, $t_1$, $q$} \;
					}
				}
			}{
				[...] \tcp{Handle other operations}
			}
			\KwRet{$ctx$} \tcp{Some ops modify the context}
		}
		
		\caption[Generation of the agendas with PCG.]{Generation of the agendas with PCG. For brevity only one case is shown in \texttt{ApplyOperation}, see the Appendix for the full list.}
		\label{alg:AgendaGeneration}
	\end{algorithm}
	
	%--------------------------------------------------------------------
	\subsubsection{Changing person focus}
	
	In our system we define one new operation that modifies the current \emph{focused} person: 
	%NURIA: REPETITIVO: %, which is an important feature introduced in our system that allows us to create an individual agenda at a time while keeping consistency with its household.
	%This operation is defined as follows:
	
	\begin{itemize}
		\item \textbf{members \{}\textit{cond1}\textbf{:} \textit{actions1} \textbf{$|$} \textit{cond2}\textbf{:} \textit{actions2} \textbf{$|$} ...\textbf{\}}
		%: this is the main operation at household-level. 
	\end{itemize}
	
	When this operation is executed, for each member of the household, the specified conditions are checked sequentially, and only the first one that evaluates as true (if any) will have its actions executed, and will set the \emph{focus} to the current household member. Conditions can be, for example, relative to the person's age, as shown in the example in Listing \ref{code:AgendaRuleFileExample} in Section~\ref{PCG_example}. This operation is somehow similar to a component split (\texttt{comp}) in CGA, which selects a subset of vertices, edges or faces of the current geometry based on conditions.
	
	%--------------------------------------------------------------------
	\subsubsection{Agenda-modifying operations}
	
	The main set of operations introduced in PCG is aimed at inserting different new actions into the agendas. Most of them take as arguments a start (\textit{$t_0$}) and end (\textit{$t_1$}) time. These operations will have an effect exclusively over the focused member of the household. Those instructions include, for example, to go to a building or to remain inside one. The full list appears in the appendix.
	
	%--------------------------------------------------------------------
	\subsubsection{Delayed execution of rules}
	
	This is a novel operation of \emph{PCG} for the agenda-generation operation, which allows the behavior decision to be delayed until simulation time:
	
	\begin{itemize}
		\item \textbf{delayedRule(${<}t_0{>},{<}t_1{>},{<}ruleName{>}$)}.
	\end{itemize}
	
	%This operation allows the behavior decision to be delayed until simulation time. 
	When an agent begins executing the agenda task created with this operation, a new instance of the rule interpreter is created, which starts at the specified rule. This instance disregards the operations that modify the agenda, and instead accepts those operations that define dynamic behavior (formal definitions at the Appendix).
	
	The major difference of this interpreter with respect to the executions performed by the initial agenda generation interpreter is that it is \textit{pausable/resumable} for some operations. This enables support for actions such as \textit{waiting a specified amount of time} (\texttt{wait}), or \textit{until the person has reached a certain place} (\texttt{goToZone}) before continuing with the execution of the subsequent rules. Therefore, rule actions are still executed sequentially, but some evaluations are delayed until the agent is prepared to act or has completed some condition. All these features only affect the time-span of the \texttt{delayedRule} entry in the person agenda: i.e. when the ending time is reached or the dynamic behavior is finished, it will proceed to its next item in the agenda. 
	%In case of early termination of the produced dynamic behavior, it will also do so.
	
	The operation pausing is carried out by attaching an execution context to the citizens when needed, and keeping track of the currently executed rule by means of a stack, which includes pairs of a rule and an index to the next successor item to execute. When a pausable operation is executed, it stores a condition function into this context. This function is periodically checked before advancing in the rule stack execution, and if it returns false (e.g. it is waiting until the waiting time has passed), no other step is taken. Otherwise, it is removed from the context and the execution is resumed at the next item in line of the rule successor.
	
	This dynamic generation of procedural behaviors offers great flexibility to our system, since rules can be triggered on the fly during the simulation, thus varying their impact depending on the overall state of the simulation. For example, a behavior can account for the availability of resources, such as free benches to sit on.

	%--------------------------------------------------------------------
	\subsubsection{Floating tasks}
	
	A key feature of PCG is the concept of \emph{floating tasks}, which are agenda entries that instead of starting and ending at a fixed time, can be dynamically scheduled to be executed when the person has an available time slot. This is achieved by the combination of two operations: 
	
	\begin{itemize}
		\item \textbf{floatingSlot(}\textit{${<}t_0{>},{<}t_1{>}$}\textbf{)}
		\item \textbf{floatingTask(}\textit{${<}duration{>},{<}ruleName{>},[{<}priority{>}]$}\textbf{)}
	\end{itemize}
	
	\textit{FloatingSlot()} creates a "free time" period in the agenda, when floating tasks can be executed. \textit{FloatingTask()} creates one of these tasks, indicating its maximum duration, and the rule to be triggered as a delayed execution rule, and optionally specifying a priority. This new task is then added to a \emph{floating tasks pool}.
	
	During the simulation, when a \emph{floating slot} is reached, the system chooses the highest-priority task from the current pool of floating tasks that has an applicable duration (in case of a tie, one of the candidates is chosen randomly). Then it is removed from the pool, and executed. In case of early termination of this task, it keeps choosing new tasks with the same criteria as long as they fit inside the remaining slot duration. When no suitable tasks are found, the person defaults to going/staying in the current location.

	%--------------------------------------------------------------------
	%--------------------------------------------------------------------
	\subsection{Procedural behavior simulation}
	
	Once the city information has been gathered and the citizens and high-level agendas are generated, the simulation can start. This simulation keeps track of a time of the day, and updates the persons' agenda accordingly. It then creates or destroys \emph{agents} as needed in the simulated world, accounting for the persons entering or exiting buildings.
	
	Notice that, in our system, an \emph{agent} is the entity representing a \emph{person} in the world, and which handles the low level A.I. and crowd behaviors such as collision avoidance or path-following to a current goal; whereas the term \emph{person} (or citizen) refers to the information describing the characteristics and relationships (i.e. age, gender, the generated agenda, the ID of the household etc.). This separation provides more flexibility and improves the scalability, as for "hidden" persons we do not need \emph{agents}, which are significantly more costly to simulate.
	
	%PArrafo modificado por completo:!!!! mirar el nuevo que está abajo
	%Albeit our system runs a sequential simulation, it also allows performing a jump to a different time. To simulate these jumps, besides skipping to a different item in each individual's agendas, the system interpolates with some randomness positions of persons along their the trajectory from the previous expected location to the current one, to provide an overall realistic initial configurations. Note that this usually will result in a differing configuration for a given time if compared against the sequential simulation with no jumps. However, it results configurations that looks plausible, which also will be eventually corrected as time passes.
	%NUEVO PARRAFO QUE SUSTITUYE AL TEXTO DE ARRIBA:
	Despite the fact that our system runs a sequential simulation, it also allows to perform jumps in time ($\nabla T$). This moves the simulation from the current time $t_i$ to $t_{i+\nabla T}$. To handle these jumps in time, the simulation skips to a different item in each individual's agendas corresponding to $t_{i+\nabla T}$, and computes and approximate new positions for each person (i.e. based on their position, velocity and known trajectory, it can estimate the position after $\delta T$). This provides a plausible configuration for the simulation to continue from $t_{i+\nabla T}$.

	\begin{algorithm}
		\SetKwFunction{FGetPositionAtNormalizedTime}{GetPositionAtNormalizedTime}
		\SetKwFunction{FInitializePersons}{InitializePersons}
		
		\SetAlgoLined
		\DontPrintSemicolon
		
		\SetKwProg{Fn}{Function}{:}{}
		\SetKw{kwAnd}{and}
		
		\Fn{\FGetPositionAtNormalizedTime{$q'$, $q$, $t$}}{
			
			\uIf{TaskType($q$) = "GoToBuilding"}{
				$p_0$ $\leftarrow$ GetEntrancePos(GetTargetBuilding($q'$)) \;
				$p_1$ $\leftarrow$ GetEntrancePos(GetTargetBuilding($q$)) \;
				\uIf{$p_0 \neq null$ \kwAnd $p_1 \neq null$}{
					$r$ $\leftarrow$ GetNavmeshPath($p_0$, $p_1$) \;
					$l$ $\leftarrow$ PathLength($r$) \;
					\KwRet{FindPathPositionAtDistance($r$, $t \cdot l$)} \;
				}
			}
			\KwRet{null}
		}
		
		\;
		
		\Fn{\FInitializePersons{/* time */ $t$}}{
			\ForEach{$p$ $\in$ persons}{
				$i$ $\leftarrow$ GetCurrentOrNextTaskIndex(agenda($p$), $t$) \;
				$q$ $\leftarrow$ GetAgendaTask($p$, $i$) \;
				$q'$ $\leftarrow$ GetAgendaTask($p$, $i - 1$) \;
				$t'$ $\leftarrow$ $(t \, - \, $StartTime($q$)$) \, / \, ($EndTime($q$)$ \, - \, $StartTime($q$)$)$ \;
				$l$ $\leftarrow$ GetPositionAtNormalizedTime($q'$, $q$, $t$) \;
				\eIf{$l$ $\neq$ null}{
					SpawnPersonInWorld($p$, $l$) \;
				}{
					$b$ $\leftarrow$ GetTargetBuilding($q$) \;
					$b$ $\leftarrow$ IsValid($b$) ? $b$ : GetHomeBuiling($p$) \;
					SetInsideBuilding($p$, $b$) \;
				}
			}
		}
		
		\caption{Strategy to initialize the persons in the simulation.}
		\label{alg:TimeJump}
	\end{algorithm}
	
	%\Otger{revisar los algoritmos: (1) se entienden? (2) falta poner alguno?. Supongo que haria falta la delayed execution, pero creo que hay demasiados detalles tecnicos de bajo nivel para ponerlo en pseudocodigo... El parrafo de abajo resume como se implementa, es suficiente?}
	
	%, despite nonetheless resulting different than if the simulation was run from an anterior time.
	
	Algorithm \ref{alg:TimeJump} shows our current strategy to initialize the positions of the agents. Note that it assumes the agenda has no gaps (in such case our system would have filled them with a default "stay-at-home" task).

	%-------------------------------------------------------------------------
	%-------------------------------------------------------------------------
	%-------------------------------------------------------------------------
	\section{Results}
	
	\newlength{\symbolwidth}
	\setlength{\symbolwidth}{0.16\linewidth}
	\begin{table}
		\centering
		\begin{tabular}{ccccc}
			\toprule
			%Symbol & label \\
			%\multirow{4}{*}{
			%\multicolumn{4}{c}{
			\includegraphics[trim={0 0 0 0},clip, width=\symbolwidth]{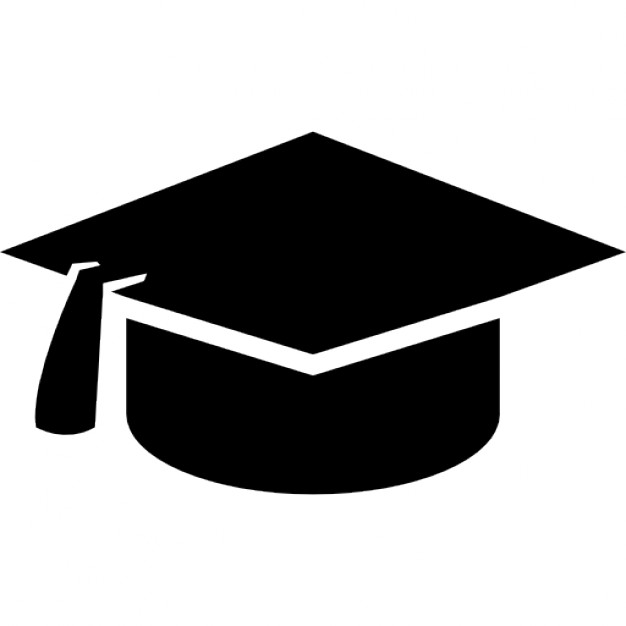} &
			\includegraphics[trim={0 0 0 0},clip, width=\symbolwidth]{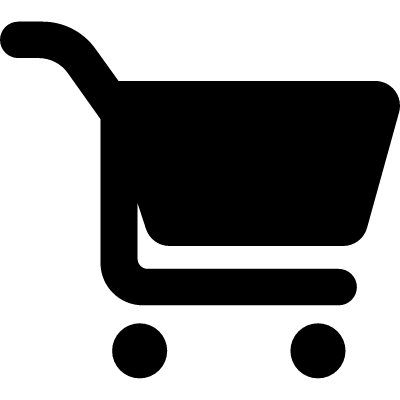} &
			\includegraphics[trim={2mm 3mm 2mm 0},clip, width=\symbolwidth]{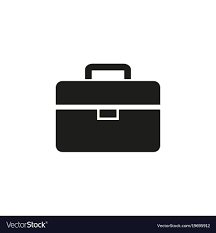} &
			\includegraphics[trim={2mm 3mm 2mm 0},clip, width=\symbolwidth]{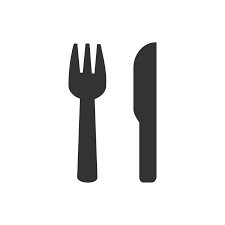} &
			\includegraphics[trim={10mm 0cm 10mm 0cm},clip, width=\symbolwidth]{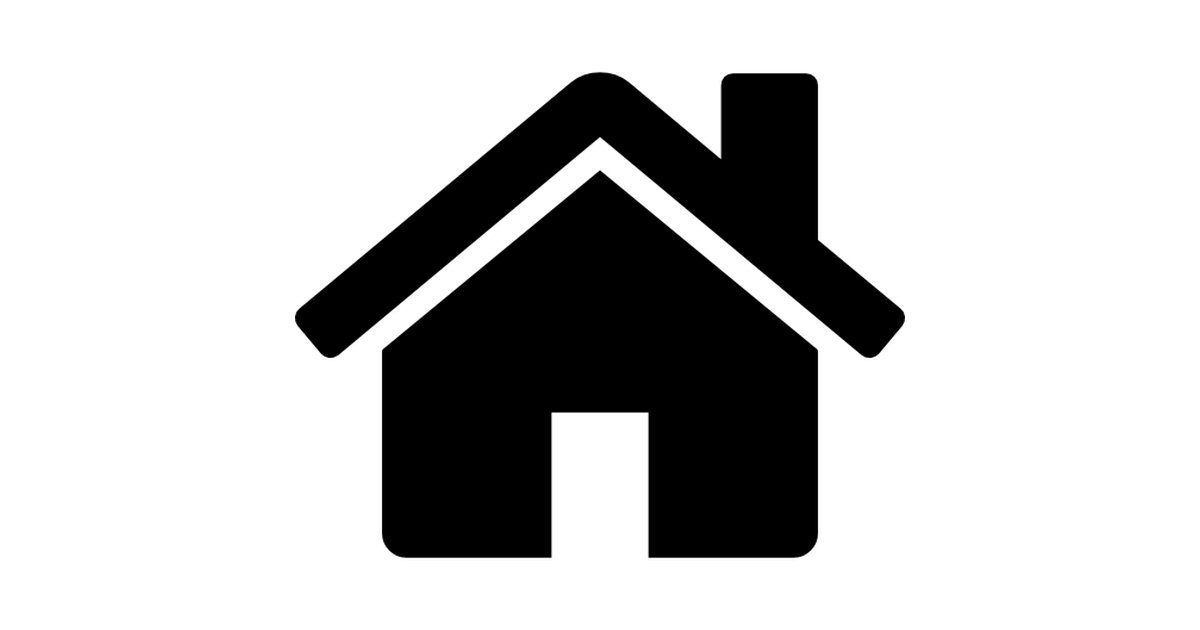} \\
			%} \\
			School & Shop & Work  & Restaurant  & House \\
			\bottomrule
		\end{tabular}
		\caption{Labels used in some of our city modeling examples.}
		\label{tbl:labels}
	\end{table}

	% \begin{figure}[h]
	% \centering
	% \includegraphics[trim={0 0 27cm 0},clip, width=0.2\linewidth]{img/figureLabels}
	% \caption{
	% Labels.}
	% \label{fig:labels}
	% \end{figure}

	\begin{figure*}[h]
		\centering
		\includegraphics[width=1\linewidth]{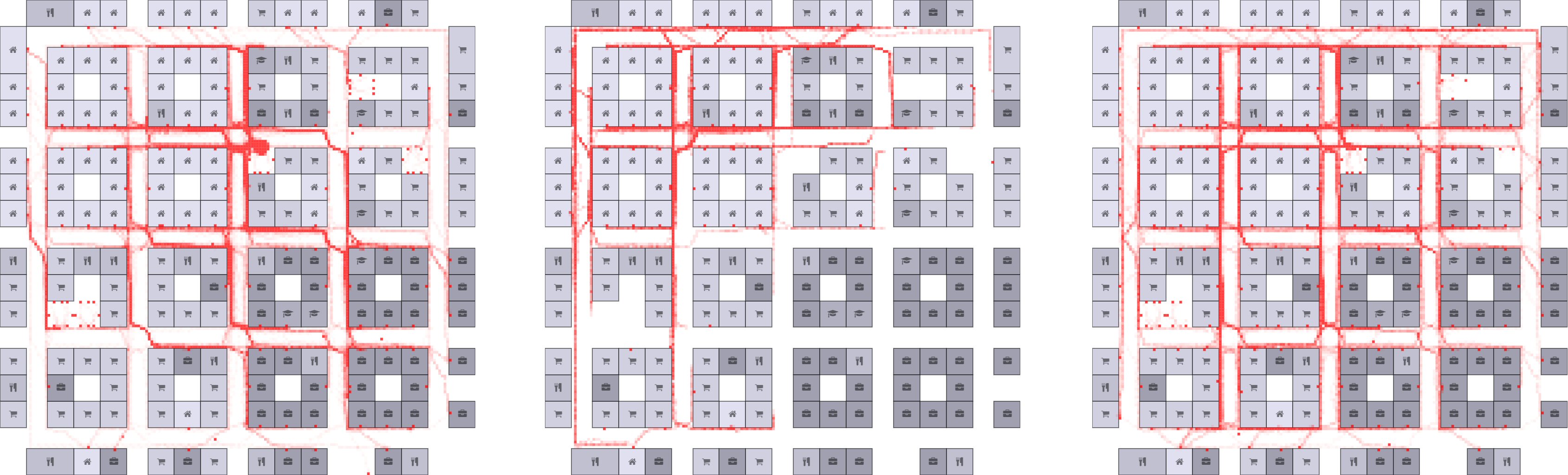}
		\caption{
			Heat-maps for 600 individuals simulated through a day, using different rule-files over a structured city (houses at top-left, workplaces bottom-right, and shops/schools on the other two quadrants). From left to right: (a) a weekday (adults go to work, and take children to school if needed); (b) a weekend where people mainly go shopping; (c). Combined rule-file adding a single rule, indicating 0.7 probability executes the weekday rule-file, and 0.3 the weekend file.}
		\label{fig:HeatmapsCombined}
	\end{figure*}
	
	\begin{figure*}[h]
		\centering
		\includegraphics[width=1\linewidth]{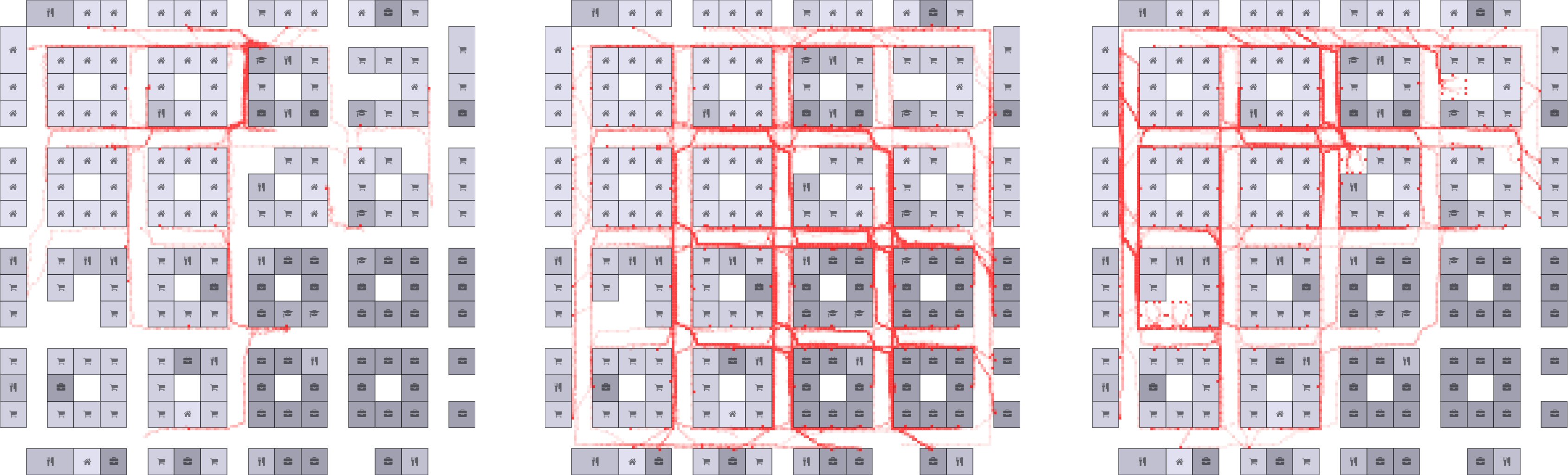}
		\caption{Heat-maps representing a simulation for different age ranges to show their agendas and paths. %The simulation used a rule-file based on the one from Listing \ref{code:BuildingRuleFileExample} but slightly extended so that some persons go shopping. 
			From left to right: children go from home to school; adults, most of which go to work taking first the kids to school, and some go shopping; elders go either shopping or to parks.}
		\label{fig:HeatmapsByAge}
	\end{figure*}
	
	%-------------------------------------------------------------------------
	%-------------------------------------------------------------------------
	%\subsection{Examples}
	%, but part of the code has been separated into a base class to keep some degree of modularity. A base class \texttt{CitizenManagerBase} contains this logic, and most of the engine-dependent code is on a class \texttt{ACitizenManager} that is itself an actor (game object) for the engine. \\
	%-------------------------------------------------------------------------
	%\subsubsection{Example of agenda-generation PCG rule set}
	
	Listing~\ref{code:AgendaRuleFileExample} in Section~\ref{PCG_example} presents a PCG rule file example used to generate agendas, which illustrates most of the supported features. The example is self-descriptive (the definition of all the functions can be found in the appendix). This simple rule file can generate households with adults and kids, with their respective tasks and agendas. Kids have to be accompanied to school at the right time and adults can then go to work at their specific working hours. Buildings are chosen based on distance or other factors. Delayed rules are used to send people to the park when they have free time in their schedule, and once in the park find a bench to sit down if there is any available. This simple example file can create an entire population for our semantically enriched city, without any need to specify individual trajectories. Users can interactively modified objects locations (add/remove/copy) and the agents will automatically make the right use of those objects (in terms of usage, times, checking for availability, etc.), or alter the type of buildings or population type, and have the crowd behavior automatically updated with consistent trajectories.
	
	To observe the resulting population whereabouts, we have created heat-maps for a population of 600 individuals (see Figure~\ref{fig:HeatmapsCombined}.) To simplify the visualization, we use a city model where we have enforced groups of building types at specific locations (See Table~\ref{tbl:labels} for the symbols used.). The left-most image shows the simulation created from the rule-file in Listing \ref{code:AgendaRuleFileExample}, which represents a weekday. The center figure shows the results of a different rule-file that defines a weekend (thus nobody goes to work/school and some people go shopping). To show the re-usability and flexibility of our system, the right-most map corresponds to the combination of different rule-files, where by adding one more rule we can combine the two previous files to create an intermediate situation. 
	
\begin{figure}[h]
		\centering
		\includegraphics[width=1\linewidth]{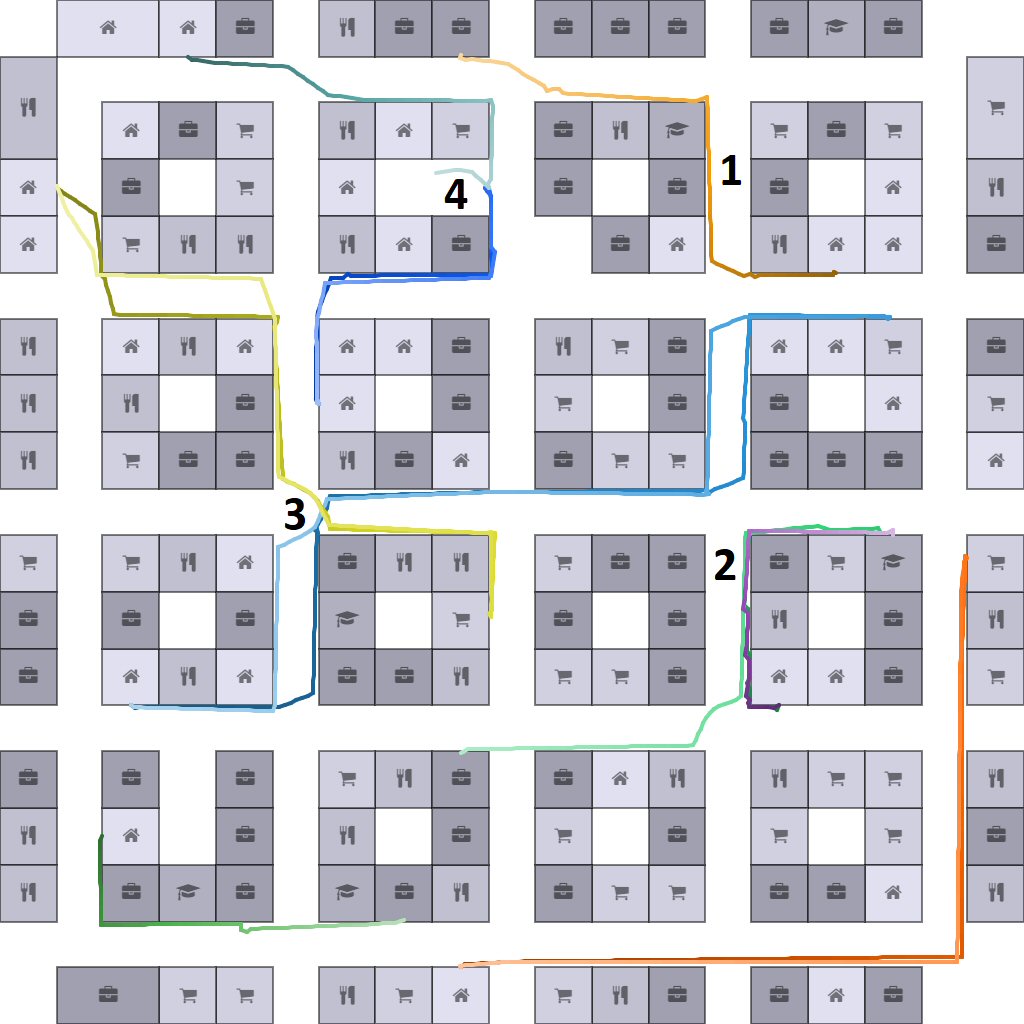}
		\caption{Example paths from a subset of individuals in a simulation that uses the rule from Figure \ref{fig:HeatmapsByAge} in a more randomly generated city. Color gradients are used to represent time, with darker meaning earlier. Some example paths: (1) simple case of an adult leaving home and going to work; (2) example of an adult (green) taking his/her child (purple) to school (with colors overlapping), and then the adult continues alone to his/her work; (3) two persons go shopping and then back home; (4) two elders head to a park; one found an empty bench and sat there, while the other one found all benches in use and went back home.}
		\label{fig:Paths}
	\end{figure}

	Similarly, figure \ref{fig:HeatmapsByAge} shows the individual behaviors based on age ranges. We can observe how the kids' heat-map on the left (from home to schools) is a subset of the adults one on the center (from home to school if they have kids, and then to work), and on the right, the heat-map for the elders (from home to parks, or groceries). Furthermore, Figure~\ref{fig:Paths} shows the paths followed by a few representative individuals. We can observe rich behaviors such as people accompanying others, something that a purely random or probability-based system would not be able to handle consistently. Also, assigning all the described behaviors for 600 individuals required writing a rule-file of about 70 lines. More importantly, this rule-file adapts to any city layout with no extra effort required by the user. %An example of this are the two layouts used in Figures \ref{fig:HeatmapsCombined} and \ref{fig:Paths}.
	
\begin{figure}[h]
		\centering
		\includegraphics[width=1\linewidth]{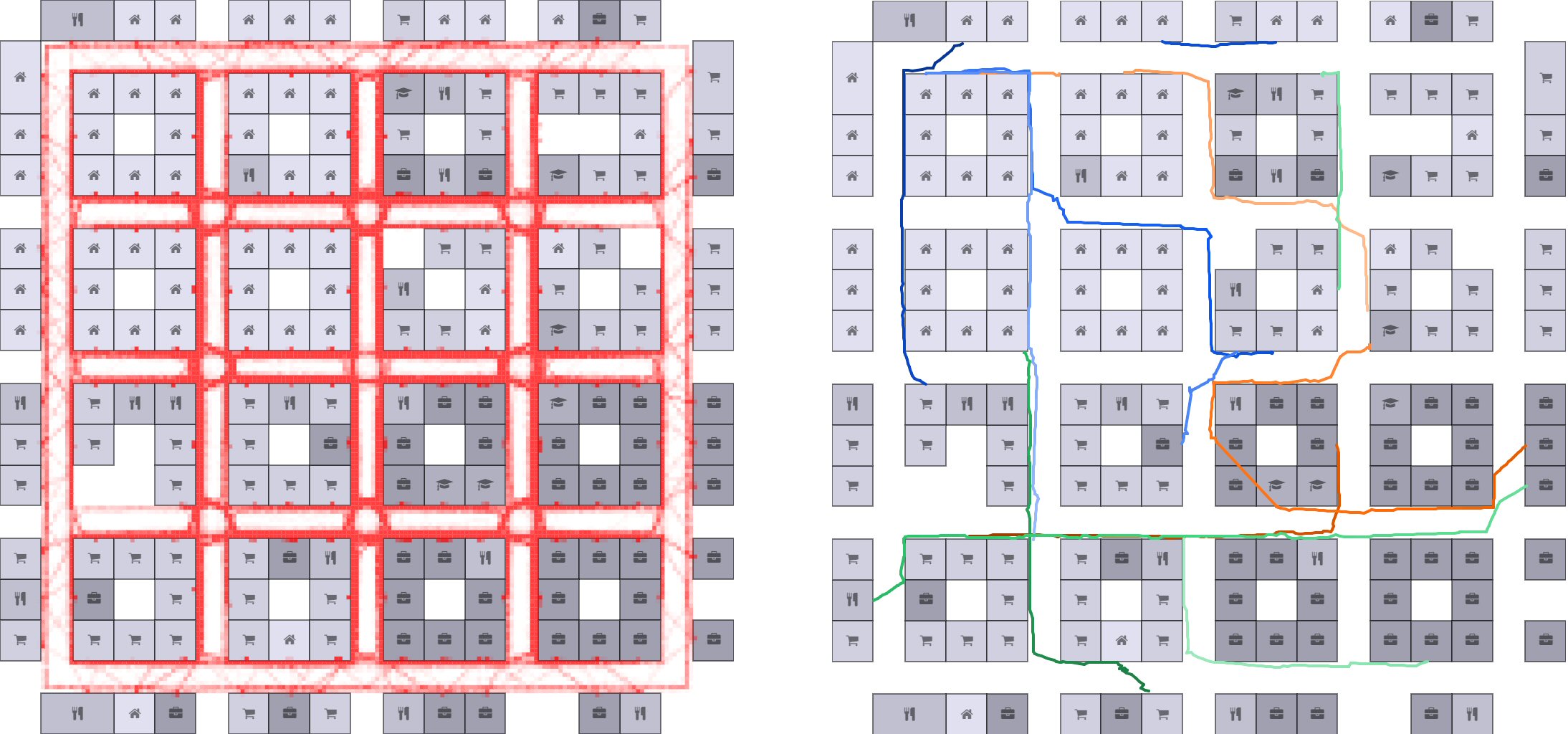}
		\caption{Heat-map and three sample paths from a fully random simulation, where individuals move between randomly chosen buildings. Notice the homogeneity of the heatmap, and the incoherent entering and exiting of buildings in the paths for a 24h simulation.}
		\label{fig:FullRandom}
	\end{figure}

	Comparing our results against a fully random simulation (see Figure~\ref{fig:FullRandom}), where individuals travel between randomly assigned buildings, We can clearly observe how our system allows for different behaviors based on factors such as age or current time, and rich behaviors such walking in group, while it also inherently prevents inconsistent cases such as individuals entering and exiting from different buildings. It is important to remark that our system can also accommodate some degree of randomness to increase variability, but, as this randomness is included at the rule-set level, always guaranteeing consistent high level behaviors that enrich the overall plausibility of the crowd.

	%-------------------------------------------------------------------------
	%-------------------------------------------------------------------------
	\subsection{Integration and User Interface}
	
	%The modules developed for this work -the procedural generation of semantically augmented cities, the Procedural Crowd Generator (\emph{PCG}) system, and the simulation- have been all integrated into the game engine Unreal Engine 4 (UE4)~\cite{UnrealEngine4}.
	
	PCG has been integrated with Unreal 4 Game Engine. Thus path finding, local movement and animations are taken care of by the engine, and our work focuses on the higher level agenda generation and simulation of consistent whereabouts for the crowd. The editor allows us to visualize and edit intermediate results, such as geometry, semantics, layout, or location of objects.

	\begin{figure}[h]
		\centering
		\includegraphics[width=1\linewidth]{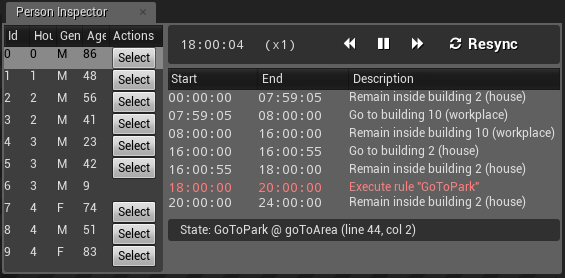}
		\caption{An example of generated agenda, seen in the inspector view implemented for the UE4 editor.}
		\label{fig:PersonInspector}
	\end{figure}
	
	%By integrating the simulation into a professional engine, we benefit from having automatic and dynamic computation of the navigation mesh, as well as having access to asset loading, citizen actors and their AI controllers, among others. The user may take advantage of the editor supplied with the engine, which we extended to integrate our framework. 
	%The editor can be used to visualize the intermediate results of the different city generation steps, as well as to run the simulation and tune behaviors. We take advantage of the editor features to allow the user to adjust the different parameters for the various steps, as well as to try and preview the results of different values for the generation of both the layout of the city, or the geometry and semantics in individual lots.

	%Whilst the generation of the city may be executed inside the editing mode, the
	
	%The simulation is performed inside the \emph{Play in Editor} mode. When this mode is started, both the population and the agendas are generated as a preprocess step. After this, a controlling system handles the control of the persons and their corresponding agents, which are created and destroyed, and assigned target locations, as required by the person's situation (inside/outside a building) and their current agenda entry.
	
	City, population and agendas are generated as a preprocess step. Then while the simulation is running, the user may also preview the agenda for each individual, to consult their personal information (age, sex, etc.) along with their schedule. %the entries in their agenda. 
	This information is shown on the Person Inspector view (see Figure \ref{fig:PersonInspector}).%that was implemented for this purpose (see figure~\ref{fig:PersonInspector}).

	%NURIA: DE MOMENTO ESTA FIGURA VA FUERA, PARA ACORTAR EL PAPER. NO APORTA NADA QUE NO SE PUEDA EXPLICAR EN PALABRAS
	%\begin{figure}[h]
	%\centering
	%\includegraphics[width=1\linewidth]{img/ResultEffectRand}
	%\caption{Comparison of the effects of adding a small offset in the start or end times. Top: $\pm$ 0, Bottom: $\pm$ 30s. Notice the presence of some regular behaviors like queues (in red).}
	%\label{fig:ResultEffectRand}
	%\end{figure}
	
	\begin{figure*}[h]
		\centering
		\includegraphics[width=0.9\linewidth]{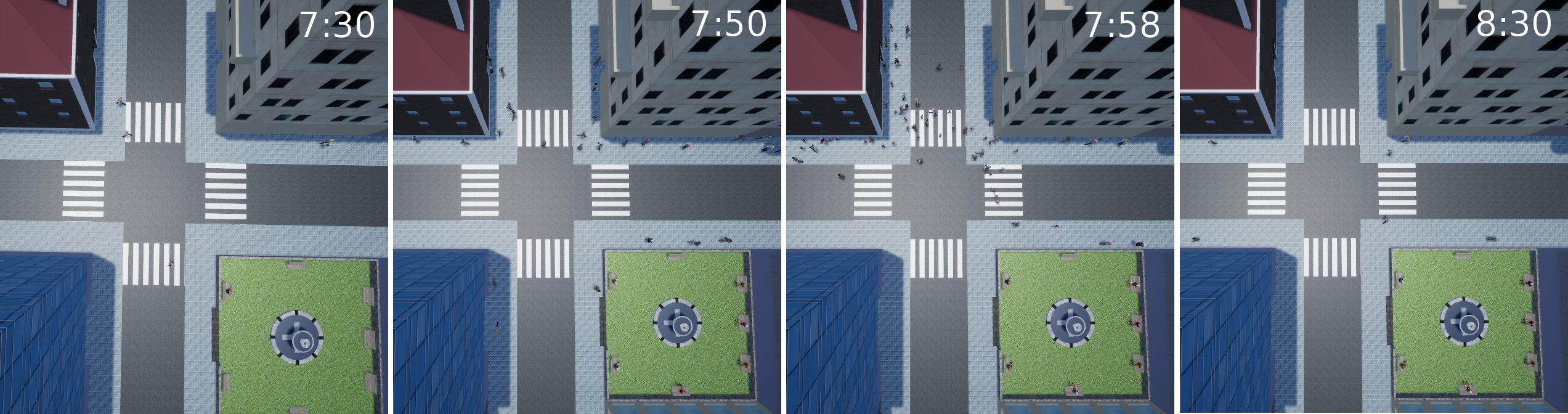}
		\caption{A simulation at different times. We can see how the amount of people in the street heading to work/schools from home changes during the morning, as well as park usage.}
		
		%At 07:30, only a few people are going to work. At 7:50 there is more people heading there (as they start at 8:00). At 7:58 still some people is going to their workplace, but also children are going (alone, when they are old enough, or accompanied) to school. Also, the park is starting to be occupied. Much later, at 10:00, there are only a few people doing errands (going to a shop), and also the benches of the park have been occupied by elder people.}
		
		\label{fig:ResultSequence}
	\end{figure*}
	
	%NURIA: ESTO COMO MUCHO LO INDICARÍA EN CONCLUSIONES Y FUTURE WORK
	%Notice that in these examples the crowd seems also very diverse in appearance, but in fact only 6 different models are used (all combinations of male/female and child/adult/elder). In the population generation step, we added a random decision for the hair color (shared in a household), and the clothing color (individual for each), which are applied by shifting respectively the lightness and hue of the corresponding texture regions. Although this appears rather random, in the future we could have the appearance also generated procedurally based on the persons' characteristics or personality. 
	
	%With figure \ref{fig:ResultEffectRand} we want to remark the potential of our method. In the top, we can observe the formation of regular behaviors (e.g., queues) in the crowd because multiple persons are using the same start time for some tasks. Yet, this can be easily solved by just adding a small offset factor to this time, achieving a more realistic scenario, with just a small change in one of the rules in the rule set. And for larger rule sets, we can repeat this at other points, thus providing every citizen a unique behavior with minimal effort.

	Figure \ref{fig:ResultSequence} displays different moments of a morning scene for some streets of a city, showing the evolution of the behaviors. In this example, we can see how the volume of people increases heading to work and children going to school as we approach 08:00 (the school and work starting hours), but there is still some movement a few hours later (e.g., some elder women are set to go to a grocery). Notice also the occupation of the park, where most elder people are set to go. Once there, if they noticed that there were no free benches anymore, they returned to their homes, as real people would do.

	%Finally, figure \ref{fig:ResultFar} shows a top view of a larger city during the simulation, in which inhabitants can be seen as tiny dots. In this example, the generated city consisted of 300 lots, and we simulated 2000 persons in it. In a middle-range PC (i5-4690 with 8GB of RAM) this simulation still ran at interactive frame-rates (50+ FPS). This is in part thanks to using an optimized engine, but nonetheless this also highlights that the actual simulation of the high-level citizens behavior has a very low footprint, the bottleneck is found on the rendering side as well as the the low-level behavior (i.e. collision avoidance and path following) required for the persons that are currently present on the map.
	
	%Our system can run in real time (50+ FPS for a city of 300 buildings and 2000 person) even in a middle range PC, since the actual simulation of the high-level citizens behavior has a very low footprint, the bottleneck is found on the rendering side as well as the the low-level behavior (i.e. collision avoidance and path following) required for the persons that are currently present on the map.
	
	%\begin{figure}[h!]
	%\centering
	%\includegraphics[width=1\linewidth]{img/ResultFar}
	%\caption{A section of a larger generated city (300 buildings), during a %simulation with 2000 persons.}
	%\label{fig:ResultFar}
	%\end{figure}
	
	We would like to emphasize the simplicity of the rule editing process, as authors have not to worry of any agenda-related conflict since the framework resolves collisions in agenda entries, and they may also use delayed-execution rules to perform actions depending on the current state of the environment. %Although at this moment the grammar is limited since it is a prototype, 
	Although limited in number of operations and functions, the current system has allowed us to evaluate its potential at the authoring level, where we could verify that the simulation system and its reusability are two of the biggest advantages when trying to populate large virtual environments. Another advantage of the system, is that as more interactable objects are included in the framework, the potential for authoring environments increases greatly. Currently, it is possible for example to copy\&paste or move benches in a park or street and then run a simulation where agents will interact correctly with them.
	
	%\Gus{Add information about interactive editing: adding a bench on the fly, and it'll work!!! Probably, an illustrating image...}
	%\Otger{lo del banc no se si es veure nomes amb una imatge, i no se com dir-ho en el text. I tal com esta ara realment no es "on-the-fly", sino parar, fer els cambis, i re-executar, veient com ara la simulacio s'adapta. Crec que quedaria millor a exemples parlar d'aixo, ja que lliga amb el parraf aquest de dalt, pero no se molt be que dir...}
	
	%\Gus{Any other "cool" feature to explain? Even if it seems trivial, it might impress the reviewers.}

	\subsection{Qualitative comparison}
	
	% Evaluation
	PCG offers a new, rule-based procedural approach to author high level consistent behaviors for all inhabitants in a crowd. Most current crowd simulation methods are not intended to give individuals a meaningful purpose, but target the crowd movement (e.g. flows, steering, local avoidance, densities or formations), without paying attention to individuals' high level behaviors. Therefore the evaluation of PCG as a simulation tool by comparison against other approaches is difficult. Of the few works that attempt to include some meaning to the individuals in a crowd, such as the works by 
	%Badler et al.~\cite{Badler2000parameterized},
	Allbeck and Badler~\cite{allbeck2002toward},
	Maim et al.~\cite{VAST:VAST07:109-116},
	Li and Allbeck~\cite{LiA11},
	Pelkey and Allbeck \cite{pelkey2014populating},
	Jorgensen et al.~\cite{jorgensen2015scheduling}, %jorgensen2014space,
	Bulbul et al.~\cite{bulbul2010populated}, or the more recent ones by 
	M. Kapadia~\cite{Kapadia:2016:MIG},
	Krontiris et al.~\cite{Krontiris16},
	and Kapadia et al.~\cite{Kapadia:2016:SCA}, it is easy to see that PCG is able to accommodate their results and provide a much larger and flexible framework for high level behavior specification. Any direct comparison with those would not only be unfeasible, but also unfair one way or another, as our goals differ.
	
	Although direct comparison is difficult, we can qualitatively compare our approach with the few works in the literature that have a goal closer to ours.
	For example, Ontology-based models \cite{Pavia2005} also aimed at generating semantic-driven crowds. In this case, generic agent profiles are defined by hand, and then assigned to individuals. The results may appear somehow similar to ours, although their approach is quite different. The main issue is the rigidity of the profile system, where specifying slightly different behaviors would require defining multiple profiles, while in our system this can be accommodated by just adding some controlled randomness at a certain rule. In addition, the ontology-based model does not support multi-agent behaviors such as going to a certain place in group, and only deals with the individual level. Our system, on the contrary can plan group behavior by creating consistent agendas automatically and thus having individuals doing things together (e.g. going to school accompanied).
	
	Another relevant work for our comparison, is the one by Kraayenbrink et al. \cite{kraayenbrink2012semantic}. They proposed a system based on satisfying agent ambitions and desires. Although this provides for a very rich simulation of agents, it runs on heuristic formulas and optimization, which limits it to small crowds of 100-200 persons. With our work we want to achieve a compromise of rich behavior and fast simulation times, enabling much larger crowds. Also, the user needs to define mathematical formulas to evaluate the ambitions, which are not intuitive, and may become difficult to balance as the amount of behaviors increases.
	
	Finally, we also wanted to compare against state-of-the-art video games, although in most cases little technical details can be found regarding their crowd implementation. This is also further obscured as developers sometimes confuse terminology by calling behaviors what are simply animations. By observation, in general we found that NPC behaviors are currently tackled in a two-level way: main characters (e.g., companions) tend to have rich behaviors defined manually using techniques such as Behavior Trees, while background characters tend to just use simple simulation methods for roaming in the world with little consistency (i.e., no actual goals) while running non-complex behaviors. With our system we try to extend the benefits of the first kind to all characters, by dealing with the main issue of requiring manual case-by-case specification.

	%-------------------------------------------------------------------------
	%-------------------------------------------------------------------------
	%-------------------------------------------------------------------------
	\section{Conclusions and future work}
	
	%% Summarize contributions:
	%The main contributions of this work are: (1) the procedural generation of semantically-augmented virtual cities with embedded information regarding land usage, zones of interest, and interactable objects; (2) an easy-to-use crowd authoring tool based on \emph{rule sets} and custom or real-world population data; (3) the procedural generation of agendas for each individual with their dynamic daily schedules; and (4) a procedural crowd simulation method. Our framework has been integrated into Unreal Engine 4~\cite{UnrealEngine4}, but designed with enough flexibility to be used within any other engine.
	We have presented a new framework for the procedural generation of semantically augmented virtual cities, which include information of land usage, zones of interest, and tagged or interactable objects. Our framework uses rule systems to generate both the city buildings geometry and semantics (an extension over the existing CGA), and the behavior of their inhabitants (using our novel PCG system), as well as custom or real-world data to generate the distribution of the population characteristics. Our results already shows these features, demonstrating not only its feasibility, but also its flexibility and design power. The success of authoring software packages like CityEngine~\cite{CityEngine}, also based on grammars and rule sets, make us confident of the practicality of the proposed method.
	
	In any traditional implementation, both the city and its population are modeled as disconnected processes that require user intervention in order to make the population react to the environment. Usually, the only automated process is the generation of a navmesh and collision information to avoid the agents to go through walls. To the best of our knowledge, this is the first procedural framework that bridges the gap between the static geometry (the city) and its dynamic population (the virtual agents), by including semantics and flexible agendas, in a fully automatic yet controllable manner.

	%As we stated earlier, there are few approaches to populate environments automatically or semi-automatically. Here we present an 
	PCG provides an authoring system that builds on the large deal of experience in procedural modeling techniques and grammar-based rule sets to provide an easy-to-use tool that allows, through some simple and logical rules, to specify complex, life-like behaviors in a population. Also, our framework allows a seamless integration with urban models in general, and procedural modeled ones in particular, resulting in a holistic tool that allows authoring a whole urban landscape and the behaviors of its inhabitants in a simple, powerful and, perhaps even more important, familiar way. It is important to remark that this technique is able to generate consistent schedules and trajectories, so it proves a very useful tool to enhance the realism of secondary characters in games, or even to provide a tool for urban planners and other stakeholders to assess the impact of new developments in the city.
	
	%% limitations and Future work
	One limitation of the current implementation is that the agendas are generated at the beginning of the simulation and remain static for the entire execution. Although we provide the means to generate dynamic and emergent behavior by using delayed-execution rules or floating tasks, it would be interesting to allow on-the-fly modification of all the agenda contents (e.g. as a result of interaction with other persons). Furthermore, relationships between tasks like precedence should be observed, as some tasks may not make sense unless another task was or will be performed.
	
	As future work, there are several directions in which the framework could be expanded. The generation of the layout of the city would be one option, in order to achieve less regular cities and increase the realism. In addition, the system could be extended in order to include vehicles, since they play an important role in the citizens' whereabouts.  
	%in order to support vehicles, as currently the generated roads are not used in the simulation.
	
	%-------------------------------------------------------------------------
	%-------------------------------------------------------------------------
	%-------------------------------------------------------------------------
	\section{Acknowledgments}
	
	This work was partially funded by the TIN2017-88515-C2-1-R and TIN2017-88515-C2-2-R projects from Ministerio de Econom\'{\i}a y Competitividad, Spain. O. Rogla has a FPU grant, funded by the Ministerio de Educaci\'{o}n, Cultura y Deporte, Spain.
	
	%-----------------------------------------------------------------------
	%-----------------------------------------------------------------------
	%-----------------------------------------------------------------------
	\bibliographystyle{ACM-Reference-Format}

	\bibliography{crowdsBib}


\begin{thebibliography}{57}

%%% the Web address.

\ifx \showCODEN    \undefined \def \showCODEN     #1{\unskip}     \fi
\ifx \showDOI      \undefined \def \showDOI       #1{#1}\fi
\ifx \showISBNx    \undefined \def \showISBNx     #1{\unskip}     \fi
\ifx \showISBNxiii \undefined \def \showISBNxiii  #1{\unskip}     \fi
\ifx \showISSN     \undefined \def \showISSN      #1{\unskip}     \fi
\ifx \showLCCN     \undefined \def \showLCCN      #1{\unskip}     \fi
\ifx \shownote     \undefined \def \shownote      #1{#1}          \fi
\ifx \showarticletitle \undefined \def \showarticletitle #1{#1}   \fi
\ifx \showURL      \undefined \def \showURL       {\relax}        \fi
% invisible to TeX
\providecommand\bibfield[2]{#2}
\providecommand\bibinfo[2]{#2}
\providecommand\natexlab[1]{#1}
\providecommand\showeprint[2][]{arXiv:#2}

\bibitem[\protect\citeauthoryear{Allbeck and Badler}{Allbeck and
  Badler}{2002}]%
        {allbeck2002toward}
\bibfield{author}{\bibinfo{person}{Jan Allbeck} {and} \bibinfo{person}{Norman
  Badler}.} \bibinfo{year}{2002}\natexlab{}.
\newblock \showarticletitle{Toward representing agent behaviors modified by
  personality and emotion}.
\newblock \bibinfo{journal}{\emph{Embodied Conversational Agents at AAMAS}}
  \bibinfo{volume}{2} (\bibinfo{year}{2002}), \bibinfo{pages}{15--19}.
\newblock
\urldef\tempurl%
\url{http://vhml.org/workshops/AAMAS/papers/allbeck.pdf}
\showURL{%
\tempurl}


\bibitem[\protect\citeauthoryear{AnonymousCity}{AnonymousCity}{2017}]%
        {Annonymous}
\bibfield{author}{\bibinfo{person}{AnonymousCity}.}
  \bibinfo{year}{2017}\natexlab{}.
\newblock \bibinfo{title}{anonymized}.
\newblock
\newblock


\bibitem[\protect\citeauthoryear{Arentze, Harry, and Timmermans}{Arentze
  et~al\mbox{.}}{2000}]%
        {Arentze_1albatross}
\bibfield{author}{\bibinfo{person}{Theo~A. Arentze}, \bibinfo{person}{Harry},
  {and} \bibinfo{person}{J.~P. Timmermans}.} \bibinfo{year}{2000}\natexlab{}.
\newblock \bibinfo{booktitle}{\emph{1 ALBATROSS – A LEARNING-BASED
  TRANSPORTATION ORIENTED SIMULATION SYSTEM}}.
\newblock \bibinfo{type}{{T}echnical {R}eport}. \bibinfo{institution}{European
  Institute of Retailing and Services Studies}.
\newblock


\bibitem[\protect\citeauthoryear{{Autodesk}}{{Autodesk}}{2016}]%
        {Maya}
\bibfield{author}{\bibinfo{person}{{Autodesk}}.}
  \bibinfo{year}{2016}\natexlab{}.
\newblock \bibinfo{title}{Maya}.
\newblock
\newblock
\urldef\tempurl%
\url{http://www.autodesk.com/products/maya/overview}
\showURL{%
\tempurl}


\bibitem[\protect\citeauthoryear{{Autodesk, Inc}}{{Autodesk, Inc}}{2016}]%
        {AutodeskCharacterGenerator}
\bibfield{author}{\bibinfo{person}{{Autodesk, Inc}}.}
  \bibinfo{year}{2016}\natexlab{}.
\newblock \bibinfo{title}{Autodesk Character Generator}.
\newblock
\newblock
\urldef\tempurl%
\url{https://charactergenerator.autodesk.com}
\showURL{%
\tempurl}


\bibitem[\protect\citeauthoryear{Badler, Bindiganavale, Palmer, Shi, and
  Schuler}{Badler et~al\mbox{.}}{2000}]%
        {Badler2000parameterized}
\bibfield{author}{\bibinfo{person}{Norman Badler}, \bibinfo{person}{Rama
  Bindiganavale}, \bibinfo{person}{Juliet Bourne~Martha Palmer},
  \bibinfo{person}{Jianping Shi}, {and} \bibinfo{person}{William Schuler}.}
  \bibinfo{year}{2000}\natexlab{}.
\newblock \showarticletitle{Parameterized action representation for virtual
  human agents}.
\newblock \bibinfo{journal}{\emph{Embodied conversational agents}}
  (\bibinfo{year}{2000}), \bibinfo{pages}{256}.
\newblock
\urldef\tempurl%
\url{http://www.cs.vu.nl/~eliens/archive/refs/par.pdf}
\showURL{%
\tempurl}


\bibitem[\protect\citeauthoryear{{Basefount}}{{Basefount}}{2016}]%
        {Maya-Miarmy}
\bibfield{author}{\bibinfo{person}{{Basefount}}.}
  \bibinfo{year}{2016}\natexlab{}.
\newblock \bibinfo{title}{Miarmy - Better Crowd Simulation for Maya}.
\newblock
\newblock
\urldef\tempurl%
\url{http://www.basefount.com/miarmy.html}
\showURL{%
\tempurl}


\bibitem[\protect\citeauthoryear{Bene\v{s}, Wilkie, and
  K\v{r}iv\'{a}nek}{Bene\v{s} et~al\mbox{.}}{2014}]%
        {Benes:2014:PMU:3071829.3071839}
\bibfield{author}{\bibinfo{person}{Jan Bene\v{s}}, \bibinfo{person}{Alexander
  Wilkie}, {and} \bibinfo{person}{Jaroslav K\v{r}iv\'{a}nek}.}
  \bibinfo{year}{2014}\natexlab{}.
\newblock \showarticletitle{Procedural Modelling of Urban Road Networks}.
\newblock \bibinfo{journal}{\emph{Comput. Graph. Forum}} \bibinfo{volume}{33},
  \bibinfo{number}{6} (\bibinfo{date}{Sept.} \bibinfo{year}{2014}),
  \bibinfo{pages}{132--142}.
\newblock
\showISSN{0167-7055}
\urldef\tempurl%
\url{https://doi.org/10.1111/cgf.12283}
\showDOI{\tempurl}


\bibitem[\protect\citeauthoryear{{Bethesda Softworks LLC}}{{Bethesda Softworks
  LLC}}{[n. d.]}]%
        {CreationKit}
\bibfield{author}{\bibinfo{person}{{Bethesda Softworks LLC}}.}
  \bibinfo{year}{[n. d.]}\natexlab{}.
\newblock \bibinfo{title}{Creation Kit}.
\newblock
\newblock
\urldef\tempurl%
\url{https://bethesda.net/community/category/131/creation-kit}
\showURL{%
\tempurl}


\bibitem[\protect\citeauthoryear{Bulbul and Dahyot}{Bulbul and Dahyot}{2016}]%
        {bulbul2010populated}
\bibfield{author}{\bibinfo{person}{Abdullah Bulbul} {and}
  \bibinfo{person}{Rozenn Dahyot}.} \bibinfo{year}{2016}\natexlab{}.
\newblock \showarticletitle{Populated Virtual Cities using Social Media}.
\newblock \bibinfo{journal}{\emph{Computer Animation and Social Agents}}
  (\bibinfo{year}{2016}).
\newblock


\bibitem[\protect\citeauthoryear{de~Paiva, Vieira, and Musse}{de~Paiva
  et~al\mbox{.}}{2005}]%
        {Pavia2005}
\bibfield{author}{\bibinfo{person}{Daniel~Costa de Paiva},
  \bibinfo{person}{Renata Vieira}, {and} \bibinfo{person}{Soraia~Raupp Musse}.}
  \bibinfo{year}{2005}\natexlab{}.
\newblock \showarticletitle{Ontology-based crowd simulation for normal life
  situations}. In \bibinfo{booktitle}{\emph{Computer Graphics International
  2005}}. IEEE, \bibinfo{pages}{221--226}.
\newblock


\bibitem[\protect\citeauthoryear{Emilien, Vimont, Cani, Poulin, and
  Benes}{Emilien et~al\mbox{.}}{2015}]%
        {emilien2015worldbrush}
\bibfield{author}{\bibinfo{person}{Arnaud Emilien}, \bibinfo{person}{Ulysse
  Vimont}, \bibinfo{person}{Marie-Paule Cani}, \bibinfo{person}{Pierre Poulin},
  {and} \bibinfo{person}{Bedrich Benes}.} \bibinfo{year}{2015}\natexlab{}.
\newblock \showarticletitle{WorldBrush: Interactive Example-based Synthesis of
  Procedural Virtual Worlds}.
\newblock \bibinfo{journal}{\emph{ACM transactions on Graphics, Proceedings of
  ACM SIGGRAPH}} \bibinfo{volume}{34}, \bibinfo{number}{4}
  (\bibinfo{year}{2015}), \bibinfo{pages}{11}.
\newblock
\urldef\tempurl%
\url{https://hal.inria.fr/hal-01147913/file/EVCPB15_Worldbrush_SIGGRAPH_2015.pdf}
\showURL{%
\tempurl}


\bibitem[\protect\citeauthoryear{{Esri}}{{Esri}}{2016}]%
        {CityEngine}
\bibfield{author}{\bibinfo{person}{{Esri}}.} \bibinfo{year}{2016}\natexlab{}.
\newblock \bibinfo{title}{Esri CityEngine}.
\newblock
\newblock
\urldef\tempurl%
\url{http://www.esri.com/software/cityengine}
\showURL{%
\tempurl}


\bibitem[\protect\citeauthoryear{Feng, Yu, Yeung, Yin, and Zhou}{Feng
  et~al\mbox{.}}{2016}]%
        {feng2016crowdlayouts}
\bibfield{author}{\bibinfo{person}{Tian Feng}, \bibinfo{person}{Lap-Fai Yu},
  \bibinfo{person}{Sai-Kit Yeung}, \bibinfo{person}{KangKang Yin}, {and}
  \bibinfo{person}{Kun Zhou}.} \bibinfo{year}{2016}\natexlab{}.
\newblock \showarticletitle{Crowd-driven Mid-scale Layout Design}.
\newblock  \bibinfo{volume}{35}, \bibinfo{number}{4} (\bibinfo{year}{2016}).
\newblock


\bibitem[\protect\citeauthoryear{Garcia-Dorado, Aliaga, and
  Ukkusuri}{Garcia-Dorado et~al\mbox{.}}{2014}]%
        {Garcia-Dorado10.1111:cgf.12329}
\bibfield{author}{\bibinfo{person}{Ignacio Garcia-Dorado},
  \bibinfo{person}{Daniel~G. Aliaga}, {and} \bibinfo{person}{Satish~V.
  Ukkusuri}.} \bibinfo{year}{2014}\natexlab{}.
\newblock \showarticletitle{{Designing Large-Scale Interactive Traffic
  Animations for Urban Modeling}}.
\newblock \bibinfo{journal}{\emph{Computer Graphics Forum}}
  (\bibinfo{year}{2014}).
\newblock
\showISSN{1467-8659}
\urldef\tempurl%
\url{https://doi.org/10.1111/cgf.12329}
\showDOI{\tempurl}


\bibitem[\protect\citeauthoryear{{Golaem}}{{Golaem}}{2016}]%
        {Maya-GOLAEM}
\bibfield{author}{\bibinfo{person}{{Golaem}}.} \bibinfo{year}{2016}\natexlab{}.
\newblock \bibinfo{title}{Golaem - Population tools for Maya}.
\newblock
\newblock
\urldef\tempurl%
\url{http://golaem.com}
\showURL{%
\tempurl}


\bibitem[\protect\citeauthoryear{Hendrikx, Meijer, Van Der~Velden, and
  Iosup}{Hendrikx et~al\mbox{.}}{2013}]%
        {SurveyProceduralGames}
\bibfield{author}{\bibinfo{person}{Mark Hendrikx}, \bibinfo{person}{Sebastiaan
  Meijer}, \bibinfo{person}{Joeri Van Der~Velden}, {and}
  \bibinfo{person}{Alexandru Iosup}.} \bibinfo{year}{2013}\natexlab{}.
\newblock \showarticletitle{Procedural Content Generation for Games: A Survey}.
\newblock \bibinfo{journal}{\emph{ACM Trans. Multimedia Comput. Commun. Appl.}}
  \bibinfo{volume}{9}, \bibinfo{number}{1}, Article \bibinfo{articleno}{1}
  (\bibinfo{date}{Feb.} \bibinfo{year}{2013}), \bibinfo{numpages}{22}~pages.
\newblock
\showISSN{1551-6857}
\urldef\tempurl%
\url{https://doi.org/10.1145/2422956.2422957}
\showDOI{\tempurl}


\bibitem[\protect\citeauthoryear{Imamura, Shirakami, Namiki, Prasertvithyakarn,
  Yokoyama, and Miyake}{Imamura et~al\mbox{.}}{2016}]%
        {Imamura:2016:FFX:2897839.2927449}
\bibfield{author}{\bibinfo{person}{Noriyuki Imamura}, \bibinfo{person}{Youji
  Shirakami}, \bibinfo{person}{Kousuke Namiki}, \bibinfo{person}{Prasert
  Prasertvithyakarn}, \bibinfo{person}{Takanori Yokoyama}, {and}
  \bibinfo{person}{Youichiro Miyake}.} \bibinfo{year}{2016}\natexlab{}.
\newblock \showarticletitle{Final Fantasy XV: Pulse and Traction of
  Characters}. In \bibinfo{booktitle}{\emph{ACM SIGGRAPH 2016 Talks}}
  \emph{(\bibinfo{series}{SIGGRAPH '16})}. \bibinfo{publisher}{ACM},
  \bibinfo{address}{New York, NY, USA}, Article \bibinfo{articleno}{47},
  \bibinfo{numpages}{2}~pages.
\newblock
\showISBNx{978-1-4503-4282-7}


\bibitem[\protect\citeauthoryear{Jordao, Charalambous, Christie, Pettr{\'e},
  and Cani}{Jordao et~al\mbox{.}}{2015}]%
        {jordao2015crowd}
\bibfield{author}{\bibinfo{person}{Kevin Jordao}, \bibinfo{person}{Panayiotis
  Charalambous}, \bibinfo{person}{Marc Christie}, \bibinfo{person}{Julien
  Pettr{\'e}}, {and} \bibinfo{person}{Marie-Paule Cani}.}
  \bibinfo{year}{2015}\natexlab{}.
\newblock \showarticletitle{Crowd art: density and flow based crowd motion
  design}. In \bibinfo{booktitle}{\emph{Motion In Games}}.
\newblock
\urldef\tempurl%
\url{http://people.rennes.inria.fr/Kevin.Jordao/ens/Doctorat/mig2015-crowdart.pdf}
\showURL{%
\tempurl}


\bibitem[\protect\citeauthoryear{Jordao, Pettr{\'e}, Christie, and Cani}{Jordao
  et~al\mbox{.}}{2014}]%
        {jordao2014crowd}
\bibfield{author}{\bibinfo{person}{Kevin Jordao}, \bibinfo{person}{Julien
  Pettr{\'e}}, \bibinfo{person}{Marc Christie}, {and} \bibinfo{person}{M-P
  Cani}.} \bibinfo{year}{2014}\natexlab{}.
\newblock \showarticletitle{Crowd sculpting: A space-time sculpting method for
  populating virtual environments}.
\newblock \bibinfo{journal}{\emph{Computer Graphics Forum}}
  \bibinfo{volume}{33}, \bibinfo{number}{2} (\bibinfo{year}{2014}),
  \bibinfo{pages}{351--360}.
\newblock
\urldef\tempurl%
\url{http://people.rennes.inria.fr/Kevin.Jordao/ens/Doctorat/crowd_sculpting.pdf}
\showURL{%
\tempurl}


\bibitem[\protect\citeauthoryear{Jorgensen}{Jorgensen}{2015}]%
        {jorgensen2015scheduling}
\bibfield{author}{\bibinfo{person}{Carl-Johan Jorgensen}.}
  \bibinfo{year}{2015}\natexlab{}.
\newblock \bibinfo{title}{Scheduling activities under spatial and temporal
  constraints to populate virtual urban environments}.
\newblock
\newblock
\urldef\tempurl%
\url{https://tel.archives-ouvertes.fr/tel-01216740/document}
\showURL{%
\tempurl}


\bibitem[\protect\citeauthoryear{Jorgensen and Lamarche}{Jorgensen and
  Lamarche}{2014}]%
        {jorgensen2014space}
\bibfield{author}{\bibinfo{person}{Carl-Johan Jorgensen} {and}
  \bibinfo{person}{Fabrice Lamarche}.} \bibinfo{year}{2014}\natexlab{}.
\newblock \bibinfo{booktitle}{\emph{Space and Time Constrained Task Scheduling
  for Crowd Simulation}}.
\newblock \bibinfo{type}{Research Report} PI 2013. \bibinfo{pages}{14} pages.
\newblock
\urldef\tempurl%
\url{https://hal.inria.fr/hal-00940570/document}
\showURL{%
\tempurl}


\bibitem[\protect\citeauthoryear{Kallmann and Thalmann}{Kallmann and
  Thalmann}{1999}]%
        {kallmann1999modeling}
\bibfield{author}{\bibinfo{person}{Marcelo Kallmann} {and}
  \bibinfo{person}{Daniel Thalmann}.} \bibinfo{year}{1999}\natexlab{}.
\newblock \showarticletitle{Modeling objects for interaction tasks}.
\newblock In \bibinfo{booktitle}{\emph{Computer Animation and
  Simulation’98}}. \bibinfo{publisher}{Springer}, \bibinfo{pages}{73--86}.
\newblock


\bibitem[\protect\citeauthoryear{Kapadia, Frey, Shoulson, Sumner, and
  Gross}{Kapadia et~al\mbox{.}}{2016a}]%
        {Kapadia:2016:SCA}
\bibfield{author}{\bibinfo{person}{Mubbasir Kapadia}, \bibinfo{person}{Seth
  Frey}, \bibinfo{person}{Alexander Shoulson}, \bibinfo{person}{Robert~W.
  Sumner}, {and} \bibinfo{person}{Markus Gross}.}
  \bibinfo{year}{2016}\natexlab{a}.
\newblock \showarticletitle{CANVAS: Computer-assisted Narrative Animation
  Synthesis}. In \bibinfo{booktitle}{\emph{Proceedings of the ACM
  SIGGRAPH/Eurographics Symposium on Computer Animation}}
  \emph{(\bibinfo{series}{SCA '16})}. \bibinfo{publisher}{Eurographics
  Association}, \bibinfo{address}{Aire-la-Ville, Switzerland, Switzerland},
  \bibinfo{pages}{199--209}.
\newblock
\showISBNx{978-3-905674-61-3}


\bibitem[\protect\citeauthoryear{Kapadia, Pelechano, Allbeck, and
  Badler}{Kapadia et~al\mbox{.}}{2015}]%
        {kapadia2015virtual}
\bibfield{author}{\bibinfo{person}{Mubbasir Kapadia}, \bibinfo{person}{Nuria
  Pelechano}, \bibinfo{person}{Jan Allbeck}, {and} \bibinfo{person}{Norm
  Badler}.} \bibinfo{year}{2015}\natexlab{}.
\newblock \showarticletitle{Virtual Crowds: Steps Toward Behavioral Realism}.
\newblock \bibinfo{journal}{\emph{Synthesis Lectures on Visual Computing:
  Computer Graphics, Animation, Computational Photography, and Imaging}}
  \bibinfo{volume}{7}, \bibinfo{number}{4} (\bibinfo{year}{2015}),
  \bibinfo{pages}{1--270}.
\newblock


\bibitem[\protect\citeauthoryear{Kapadia, Shoulson, Steimer, Oberholzer,
  Sumner, and Gross}{Kapadia et~al\mbox{.}}{2016b}]%
        {Kapadia:2016:MIG}
\bibfield{author}{\bibinfo{person}{Mubbasir Kapadia},
  \bibinfo{person}{Alexander Shoulson}, \bibinfo{person}{Cyril Steimer},
  \bibinfo{person}{Samuel Oberholzer}, \bibinfo{person}{Robert~W. Sumner},
  {and} \bibinfo{person}{Markus Gross}.} \bibinfo{year}{2016}\natexlab{b}.
\newblock \showarticletitle{An Event-centric Approach to Authoring Stories in
  Crowds}. In \bibinfo{booktitle}{\emph{Proceedings of the 9th International
  Conference on Motion in Games}} \emph{(\bibinfo{series}{MIG '16})}.
  \bibinfo{publisher}{ACM}, \bibinfo{address}{New York, NY, USA},
  \bibinfo{pages}{15--24}.
\newblock
\showISBNx{978-1-4503-4592-7}
\urldef\tempurl%
\url{https://doi.org/10.1145/2994258.2994265}
\showDOI{\tempurl}


\bibitem[\protect\citeauthoryear{Katoshevski, Katoshevski, Arentze, and
  Timmermans}{Katoshevski et~al\mbox{.}}{2014}]%
        {JTLU333}
\bibfield{author}{\bibinfo{person}{Rachel Katoshevski}, \bibinfo{person}{David
  Katoshevski}, \bibinfo{person}{Theo Arentze}, {and} \bibinfo{person}{Harry
  Timmermans}.} \bibinfo{year}{2014}\natexlab{}.
\newblock \showarticletitle{A multi-agent planning support system for assessing
  the role of transportation and environmental objectives in urban planning}.
\newblock \bibinfo{journal}{\emph{Journal of Transport and Land Use}}
  \bibinfo{volume}{7}, \bibinfo{number}{1} (\bibinfo{year}{2014}),
  \bibinfo{pages}{29--42}.
\newblock
\showISSN{1938-7849}


\bibitem[\protect\citeauthoryear{Katoshevski-Cavari, Arentze, and
  Timmermans}{Katoshevski-Cavari et~al\mbox{.}}{2011}]%
        {KatoshevskiCavari2011131}
\bibfield{author}{\bibinfo{person}{Rachel Katoshevski-Cavari},
  \bibinfo{person}{Theo~A. Arentze}, {and} \bibinfo{person}{Harry~J.P.
  Timmermans}.} \bibinfo{year}{2011}\natexlab{}.
\newblock \showarticletitle{Sustainable City-Plan Based on Planning Algorithm,
  Planner Heuristics and Transportation Aspects}.
\newblock \bibinfo{journal}{\emph{Procedia - Social and Behavioral Sciences}}
  \bibinfo{volume}{20} (\bibinfo{year}{2011}), \bibinfo{pages}{131 -- 139}.
\newblock
\showISSN{1877-0428}


\bibitem[\protect\citeauthoryear{Kim, Seol, Kwon, and Lee}{Kim
  et~al\mbox{.}}{2014}]%
        {kim2014interactive}
\bibfield{author}{\bibinfo{person}{Jongmin Kim}, \bibinfo{person}{Yeongho
  Seol}, \bibinfo{person}{Taesoo Kwon}, {and} \bibinfo{person}{Jehee Lee}.}
  \bibinfo{year}{2014}\natexlab{}.
\newblock \showarticletitle{Interactive manipulation of large-scale crowd
  animation}.
\newblock \bibinfo{journal}{\emph{ACM Transactions on Graphics (TOG)}}
  \bibinfo{volume}{33}, \bibinfo{number}{4} (\bibinfo{year}{2014}),
  \bibinfo{pages}{83}.
\newblock
\urldef\tempurl%
\url{http://mrl.snu.ac.kr/research/ProjectCrowdEditing/crowd_editing_preprint.pdf}
\showURL{%
\tempurl}


\bibitem[\protect\citeauthoryear{Kraayenbrink, Kessing, Tutenel, de~Haan,
  Marson, Musse, and Bidarra}{Kraayenbrink et~al\mbox{.}}{2012}]%
        {kraayenbrink2012semantic}
\bibfield{author}{\bibinfo{person}{Nick Kraayenbrink}, \bibinfo{person}{Jassin
  Kessing}, \bibinfo{person}{Tim Tutenel}, \bibinfo{person}{Gerwin de Haan},
  \bibinfo{person}{Fernando Marson}, \bibinfo{person}{Soraia~R Musse}, {and}
  \bibinfo{person}{Rafael Bidarra}.} \bibinfo{year}{2012}\natexlab{}.
\newblock \showarticletitle{Semantic crowds: reusable population for virtual
  worlds}.
\newblock \bibinfo{journal}{\emph{Procedia Computer Science}}
  \bibinfo{volume}{15} (\bibinfo{year}{2012}), \bibinfo{pages}{122--139}.
\newblock


\bibitem[\protect\citeauthoryear{Krontiris, Bekris, and Kapadia}{Krontiris
  et~al\mbox{.}}{2016}]%
        {Krontiris16}
\bibfield{author}{\bibinfo{person}{A. Krontiris}, \bibinfo{person}{K.~E.
  Bekris}, {and} \bibinfo{person}{M. Kapadia}.}
  \bibinfo{year}{2016}\natexlab{}.
\newblock \showarticletitle{ACUMEN: Activity-Centric Crowd Authoring Using
  Influence Maps}. In \bibinfo{booktitle}{\emph{29th International Conference
  on Computer Animation and Social Agents (CASA)}}. \bibinfo{address}{Geneva,
  Switzerland}.
\newblock
\urldef\tempurl%
\url{https://www.cs.rutgers.edu/~kb572/pubs/acumen_casa_2016.pdf}
\showURL{%
\tempurl}


\bibitem[\protect\citeauthoryear{Li and Allbeck}{Li and Allbeck}{2011}]%
        {LiA11}
\bibfield{author}{\bibinfo{person}{Weizi~Philip Li} {and}
  \bibinfo{person}{Jan~M. Allbeck}.} \bibinfo{year}{2011}\natexlab{}.
\newblock \showarticletitle{Populations with Purpose.}. In
  \bibinfo{booktitle}{\emph{MIG}} \emph{(\bibinfo{series}{Lecture Notes in
  Computer Science})}, \bibfield{editor}{\bibinfo{person}{Jan~M. Allbeck} {and}
  \bibinfo{person}{Petros Faloutsos}} (Eds.), Vol.~\bibinfo{volume}{7060}.
  \bibinfo{publisher}{Springer}, \bibinfo{pages}{132--143}.
\newblock
\showISBNx{978-3-642-25089-7}
\urldef\tempurl%
\url{http://dblp.uni-trier.de/db/conf/mig/mig2011.html#LiA11}
\showURL{%
\tempurl}


\bibitem[\protect\citeauthoryear{Maim, Haegler, Yersin, Mueller, Thalmann, and
  Gool}{Maim et~al\mbox{.}}{2007}]%
        {VAST:VAST07:109-116}
\bibfield{author}{\bibinfo{person}{Jonathan Maim}, \bibinfo{person}{Simon
  Haegler}, \bibinfo{person}{Barbara Yersin}, \bibinfo{person}{Pascal Mueller},
  \bibinfo{person}{Daniel Thalmann}, {and} \bibinfo{person}{Luc~Van Gool}.}
  \bibinfo{year}{2007}\natexlab{}.
\newblock \showarticletitle{{Populating Ancient Pompeii with Crowds of Virtual
  Romans}}. In \bibinfo{booktitle}{\emph{VAST: International Symposium on
  Virtual Reality, Archaeology and Intelligent Cultural Heritage}},
  \bibfield{editor}{\bibinfo{person}{D.~Arnold},
  \bibinfo{person}{F.~Niccolucci}, {and} \bibinfo{person}{A.~Chalmers}} (Eds.).
  \bibinfo{publisher}{The Eurographics Association}.
\newblock
\showISBNx{978-3-905674-01-9}
\showISSN{1811-864X}
\urldef\tempurl%
\url{https://doi.org/10.2312/VAST/VAST07/109-116}
\showDOI{\tempurl}


\bibitem[\protect\citeauthoryear{{MakeHuman}}{{MakeHuman}}{2016}]%
        {MakeHuman}
\bibfield{author}{\bibinfo{person}{{MakeHuman}}.}
  \bibinfo{year}{2016}\natexlab{}.
\newblock \bibinfo{title}{MakeHuman}.
\newblock
\newblock
\urldef\tempurl%
\url{http://www.makehuman.org/}
\showURL{%
\tempurl}


\bibitem[\protect\citeauthoryear{{Mixamo}}{{Mixamo}}{2016}]%
        {Mixamo}
\bibfield{author}{\bibinfo{person}{{Mixamo}}.} \bibinfo{year}{2016}\natexlab{}.
\newblock \bibinfo{title}{Mixamo}.
\newblock
\newblock
\urldef\tempurl%
\url{https://www.mixamo.com/}
\showURL{%
\tempurl}


\bibitem[\protect\citeauthoryear{M{\"u}ller, Wonka, Haegler, Ulmer, and
  Van~Gool}{M{\"u}ller et~al\mbox{.}}{2006}]%
        {muller2006procedural}
\bibfield{author}{\bibinfo{person}{Pascal M{\"u}ller}, \bibinfo{person}{Peter
  Wonka}, \bibinfo{person}{Simon Haegler}, \bibinfo{person}{Andreas Ulmer},
  {and} \bibinfo{person}{Luc Van~Gool}.} \bibinfo{year}{2006}\natexlab{}.
\newblock \showarticletitle{Procedural modeling of buildings}.
\newblock \bibinfo{journal}{\emph{Acm Transactions On Graphics (Tog)}}
  \bibinfo{volume}{25}, \bibinfo{number}{3} (\bibinfo{year}{2006}),
  \bibinfo{pages}{614--623}.
\newblock
\urldef\tempurl%
\url{http://homes.esat.kuleuven.be/~konijn/publications/2006/SIGGRAPH-PM-06.pdf}
\showURL{%
\tempurl}


\bibitem[\protect\citeauthoryear{Parish and M{\"{u}}ller}{Parish and
  M{\"{u}}ller}{2001}]%
        {Parish01}
\bibfield{author}{\bibinfo{person}{Yoav I~H Parish} {and}
  \bibinfo{person}{Pascal M{\"{u}}ller}.} \bibinfo{year}{2001}\natexlab{}.
\newblock \showarticletitle{{Procedural modeling of cities}}. In
  \bibinfo{booktitle}{\emph{Proceedings of the 28th annual conference on
  Computer graphics and interactive techniques - SIGGRAPH '01}}
  \emph{(\bibinfo{series}{SIGGRAPH '01})}. \bibinfo{publisher}{ACM Press},
  \bibinfo{pages}{301--308}.
\newblock


\bibitem[\protect\citeauthoryear{Patil, Van~den Berg, Curtis, Lin, and
  Manocha}{Patil et~al\mbox{.}}{2011}]%
        {patil2011directing}
\bibfield{author}{\bibinfo{person}{Sachin Patil}, \bibinfo{person}{Jur Van~den
  Berg}, \bibinfo{person}{Sean Curtis}, \bibinfo{person}{Ming~C Lin}, {and}
  \bibinfo{person}{Dinesh Manocha}.} \bibinfo{year}{2011}\natexlab{}.
\newblock \showarticletitle{Directing crowd simulations using navigation
  fields}.
\newblock \bibinfo{journal}{\emph{Visualization and Computer Graphics, IEEE
  Transactions on}} \bibinfo{volume}{17}, \bibinfo{number}{2}
  (\bibinfo{year}{2011}), \bibinfo{pages}{244--254}.
\newblock
\urldef\tempurl%
\url{http://gamma.cs.unc.edu/DCrowd/paper.pdf}
\showURL{%
\tempurl}


\bibitem[\protect\citeauthoryear{Pelkey and Allbeck}{Pelkey and
  Allbeck}{2014}]%
        {pelkey2014populating}
\bibfield{author}{\bibinfo{person}{Cameron~D Pelkey} {and}
  \bibinfo{person}{Jan~M Allbeck}.} \bibinfo{year}{2014}\natexlab{}.
\newblock \showarticletitle{Populating semantic virtual environments}.
\newblock \bibinfo{journal}{\emph{Computer Animation and Virtual Worlds}}
  \bibinfo{volume}{25}, \bibinfo{number}{3-4} (\bibinfo{year}{2014}),
  \bibinfo{pages}{403--410}.
\newblock
\urldef\tempurl%
\url{http://cs.gmu.edu/~gaia/publications/Semantics-CASA-2014.pdf}
\showURL{%
\tempurl}


\bibitem[\protect\citeauthoryear{Peng, Yang, Bao, Fink, Yan, Wonka, and
  Mitra}{Peng et~al\mbox{.}}{2016}]%
        {peng2016funcNetwork}
\bibfield{author}{\bibinfo{person}{Chi-Han Peng}, \bibinfo{person}{Yong-Liang
  Yang}, \bibinfo{person}{Fan Bao}, \bibinfo{person}{Daniel Fink},
  \bibinfo{person}{Dong-Ming Yan}, \bibinfo{person}{Peter Wonka}, {and}
  \bibinfo{person}{N.~J. Mitra}.} \bibinfo{year}{2016}\natexlab{}.
\newblock \showarticletitle{Computational Network Design from Functional
  Specifications}.
\newblock \bibinfo{journal}{\emph{ACM Transactions on Graphics (Proceedings of
  SIGGRAPH 2016)}}  \bibinfo{volume}{35} (\bibinfo{year}{2016}).
\newblock
Issue 4.


\bibitem[\protect\citeauthoryear{Rasouli and Timmermans}{Rasouli and
  Timmermans}{2014}]%
        {Rasouli201479}
\bibfield{author}{\bibinfo{person}{Soora Rasouli} {and} \bibinfo{person}{Harry
  Timmermans}.} \bibinfo{year}{2014}\natexlab{}.
\newblock \showarticletitle{Applications of theories and models of choice and
  decision-making under conditions of uncertainty in travel behavior research}.
\newblock \bibinfo{journal}{\emph{Travel Behaviour and Society}}
  \bibinfo{volume}{1}, \bibinfo{number}{3} (\bibinfo{year}{2014}),
  \bibinfo{pages}{79 -- 90}.
\newblock
\showISSN{2214-367X}
\urldef\tempurl%
\url{https://doi.org/10.1016/j.tbs.2013.12.001}
\showDOI{\tempurl}


\bibitem[\protect\citeauthoryear{Schwarz and Wonka}{Schwarz and Wonka}{2015}]%
        {schwarz2015practical}
\bibfield{author}{\bibinfo{person}{Michael Schwarz} {and}
  \bibinfo{person}{Peter Wonka}.} \bibinfo{year}{2015}\natexlab{}.
\newblock \showarticletitle{Practical grammar-based procedural modeling of
  architecture: SIGGRAPH Asia 2015 course notes}. In
  \bibinfo{booktitle}{\emph{SIGGRAPH Asia 2015 Courses}}. ACM,
  \bibinfo{pages}{13}.
\newblock
\urldef\tempurl%
\url{http://research.michael-schwarz.com/publ/files/procmodcourse-siga15.pdf}
\showURL{%
\tempurl}


\bibitem[\protect\citeauthoryear{Torrens}{Torrens}{2007}]%
        {Torrens07}
\bibfield{author}{\bibinfo{person}{D.~P.~M. Torrens}.}
  \bibinfo{year}{2007}\natexlab{}.
\newblock \showarticletitle{Behavioral Intelligence for Geospatial Agents in
  Urban Environments}. In \bibinfo{booktitle}{\emph{Intelligent Agent
  Technology, 2007. IAT '07. IEEE/WIC/ACM International Conference on}}.
  \bibinfo{pages}{63--66}.
\newblock
\urldef\tempurl%
\url{https://doi.org/10.1109/IAT.2007.45}
\showDOI{\tempurl}


\bibitem[\protect\citeauthoryear{Torrens, Li, and Griffin}{Torrens
  et~al\mbox{.}}{2011}]%
        {TGIS:TGIS1261}
\bibfield{author}{\bibinfo{person}{Paul Torrens}, \bibinfo{person}{Xun Li},
  {and} \bibinfo{person}{William~A. Griffin}.} \bibinfo{year}{2011}\natexlab{}.
\newblock \showarticletitle{Building Agent-Based Walking Models by
  Machine-Learning on Diverse Databases of Space-Time Trajectory Samples}.
\newblock \bibinfo{journal}{\emph{Transactions in GIS}}  \bibinfo{volume}{15}
  (\bibinfo{year}{2011}), \bibinfo{pages}{67--94}.
\newblock
\showISSN{1467-9671}
\urldef\tempurl%
\url{https://doi.org/10.1111/j.1467-9671.2011.01261.x}
\showDOI{\tempurl}


\bibitem[\protect\citeauthoryear{Torrens, Nara, Li, Zhu, Griffin, and
  Brown}{Torrens et~al\mbox{.}}{2012}]%
        {Torrens20121}
\bibfield{author}{\bibinfo{person}{Paul~M. Torrens}, \bibinfo{person}{Atsushi
  Nara}, \bibinfo{person}{Xun Li}, \bibinfo{person}{Haojie Zhu},
  \bibinfo{person}{William~A. Griffin}, {and} \bibinfo{person}{Scott~B.
  Brown}.} \bibinfo{year}{2012}\natexlab{}.
\newblock \showarticletitle{An extensible simulation environment and movement
  metrics for testing walking behavior in agent-based models}.
\newblock \bibinfo{journal}{\emph{Computers, Environment and Urban Systems}}
  \bibinfo{volume}{36}, \bibinfo{number}{1} (\bibinfo{year}{2012}),
  \bibinfo{pages}{1 -- 17}.
\newblock
\showISSN{0198-9715}
\urldef\tempurl%
\url{https://doi.org/10.1016/j.compenvurbsys.2011.07.005}
\showDOI{\tempurl}


\bibitem[\protect\citeauthoryear{(Ubisoft)}{(Ubisoft)}{2011}]%
        {AssassinsCreedBroderhood}
\bibfield{author}{\bibinfo{person}{Aleissia Laidacker Nicolas~Barbeau
  (Ubisoft)}.} \bibinfo{year}{2011}\natexlab{}.
\newblock \showarticletitle{Living Crowds AI \& Animation in: Assassin’s
  Creed: Brotherhood}. In \bibinfo{booktitle}{\emph{Game Developers
  Conference}}.
\newblock


\bibitem[\protect\citeauthoryear{(Ubisoft)}{(Ubisoft)}{2015a}]%
        {AssassinsCreedUnity2}
\bibfield{author}{\bibinfo{person}{Christine~Blondeau (Ubisoft)}.}
  \bibinfo{year}{2015}\natexlab{a}.
\newblock \showarticletitle{Postmortem: Designing systemic crowd events on
  Assassin’s Creed Unity}. In \bibinfo{booktitle}{\emph{Game Developers
  Conference}}.
\newblock


\bibitem[\protect\citeauthoryear{(Ubisoft)}{(Ubisoft)}{2015b}]%
        {AssassinsCreedUnity1}
\bibfield{author}{\bibinfo{person}{Francois~Cournoyer (Ubisoft)}.}
  \bibinfo{year}{2015}\natexlab{b}.
\newblock \showarticletitle{Massive Crowd on Assassin's Creed Unity: AI
  Recycling}. In \bibinfo{booktitle}{\emph{Game Developers Conference}}.
\newblock


\bibitem[\protect\citeauthoryear{(Ubisoft)}{(Ubisoft)}{2012}]%
        {Hitman}
\bibfield{author}{\bibinfo{person}{Kasper~Fauerby (Ubisoft)}.}
  \bibinfo{year}{2012}\natexlab{}.
\newblock \showarticletitle{Crowds in Hitman: Absolution}. In
  \bibinfo{booktitle}{\emph{Game Developers Conference}}.
\newblock


\bibitem[\protect\citeauthoryear{(Ubisoft)}{(Ubisoft)}{2017}]%
        {WatchDogs2}
\bibfield{author}{\bibinfo{person}{Roxanne Blouin-Payer (Ubisoft)}.}
  \bibinfo{year}{2017}\natexlab{}.
\newblock \showarticletitle{Helping It All Emerge: Managing Crowd AI in 'Watch
  Dogs 2'}. In \bibinfo{booktitle}{\emph{Game Developers Conference}}.
\newblock


\bibitem[\protect\citeauthoryear{Ulicny, Ciechomski, and Thalmann}{Ulicny
  et~al\mbox{.}}{2004}]%
        {Ulicny04}
\bibfield{author}{\bibinfo{person}{Branislav Ulicny}, \bibinfo{person}{Pablo
  de~Heras Ciechomski}, {and} \bibinfo{person}{Daniel Thalmann}.}
  \bibinfo{year}{2004}\natexlab{}.
\newblock \showarticletitle{{Crowdbrush: Interactive Authoring of Real-time
  Crowd Scenes}}. In \bibinfo{booktitle}{\emph{Symposium on Computer
  Animation}}, \bibfield{editor}{\bibinfo{person}{R.~Boulic} {and}
  \bibinfo{person}{D.~K. Pai}} (Eds.). \bibinfo{publisher}{The Eurographics
  Association}.
\newblock
\showISBNx{3-905673-14-2}
\showISSN{1727-5288}
\urldef\tempurl%
\url{https://doi.org/10.2312/SCA/SCA04/243-252}
\showDOI{\tempurl}


\bibitem[\protect\citeauthoryear{{Unity Technologies}}{{Unity
  Technologies}}{2017}]%
        {MatSim}
\bibfield{author}{\bibinfo{person}{{Unity Technologies}}.}
  \bibinfo{year}{2017}\natexlab{}.
\newblock \bibinfo{title}{MATSim - large-scale agent-based transport
  simulations.}
\newblock
\newblock
\urldef\tempurl%
\url{http://www.matsim.org/}
\showURL{%
\tempurl}


\bibitem[\protect\citeauthoryear{Vanegas, Aliaga, Wonka, M{\"u}ller, Waddell,
  and Watson}{Vanegas et~al\mbox{.}}{2010}]%
        {Vanegas09}
\bibfield{author}{\bibinfo{person}{Carlos~A. Vanegas},
  \bibinfo{person}{Daniel~G. Aliaga}, \bibinfo{person}{Peter Wonka},
  \bibinfo{person}{Pascal M{\"u}ller}, \bibinfo{person}{Paul Waddell}, {and}
  \bibinfo{person}{Benjamin Watson}.} \bibinfo{year}{2010}\natexlab{}.
\newblock \showarticletitle{Modelling the Appearance and Behaviour of Urban
  Spaces}.
\newblock \bibinfo{journal}{\emph{Comput. Graph. Forum}} \bibinfo{volume}{29},
  \bibinfo{number}{1} (\bibinfo{year}{2010}), \bibinfo{pages}{25--42}.
\newblock


\bibitem[\protect\citeauthoryear{Watson, M\"{u}ller, Veryovka, Fuller, Wonka,
  and Sexton}{Watson et~al\mbox{.}}{2008}]%
        {Watson08}
\bibfield{author}{\bibinfo{person}{Benjamin Watson}, \bibinfo{person}{Pascal
  M\"{u}ller}, \bibinfo{person}{Oleg Veryovka}, \bibinfo{person}{Andy Fuller},
  \bibinfo{person}{Peter Wonka}, {and} \bibinfo{person}{Chris Sexton}.}
  \bibinfo{year}{2008}\natexlab{}.
\newblock \showarticletitle{Procedural Urban Modeling in Practice}.
\newblock \bibinfo{journal}{\emph{IEEE Computer Graphics and Applications}}
  \bibinfo{volume}{28} (\bibinfo{year}{2008}), \bibinfo{pages}{18--26}.
\newblock
\showISSN{0272-1716}


\bibitem[\protect\citeauthoryear{Wonka, Wimmer, Sillion, and Ribarsky}{Wonka
  et~al\mbox{.}}{2003}]%
        {Wonka2003}
\bibfield{author}{\bibinfo{person}{Peter Wonka}, \bibinfo{person}{Michael
  Wimmer}, \bibinfo{person}{Fran{\c c}ois Sillion}, {and}
  \bibinfo{person}{William Ribarsky}.} \bibinfo{year}{2003}\natexlab{}.
\newblock \showarticletitle{Instant Architecture}.
\newblock \bibinfo{journal}{\emph{ACM Transaction on Graphics}}
  \bibinfo{volume}{22}, \bibinfo{number}{3} (\bibinfo{date}{July}
  \bibinfo{year}{2003}), \bibinfo{pages}{669--677}.
\newblock
\newblock
\shownote{Proceedings ACM SIGGRAPH 2003.}


\bibitem[\protect\citeauthoryear{Yersin, Ma\"{\i}m, Pettr{\'e}, and
  Thalmann}{Yersin et~al\mbox{.}}{2009}]%
        {Yersin:2009:CPP:1507149.1507184}
\bibfield{author}{\bibinfo{person}{Barbara Yersin}, \bibinfo{person}{Jonathan
  Ma\"{\i}m}, \bibinfo{person}{Julien Pettr{\'e}}, {and}
  \bibinfo{person}{Daniel Thalmann}.} \bibinfo{year}{2009}\natexlab{}.
\newblock \showarticletitle{Crowd Patches: Populating Large-scale Virtual
  Environments for Real-time Applications}. In
  \bibinfo{booktitle}{\emph{Proceedings of the 2009 Symposium on Interactive 3D
  Graphics and Games}} \emph{(\bibinfo{series}{I3D '09})}.
  \bibinfo{publisher}{ACM}, \bibinfo{address}{New York, NY, USA},
  \bibinfo{pages}{207--214}.
\newblock
\showISBNx{978-1-60558-429-4}
\urldef\tempurl%
\url{https://doi.org/10.1145/1507149.1507184}
\showDOI{\tempurl}


\bibitem[\protect\citeauthoryear{Zou, Torrens, Ghanem, and Kevrekidis}{Zou
  et~al\mbox{.}}{2012}]%
        {Zou12}
\bibfield{author}{\bibinfo{person}{Yu Zou}, \bibinfo{person}{Paul~M. Torrens},
  \bibinfo{person}{Roger~G. Ghanem}, {and} \bibinfo{person}{Ioannis~G.
  Kevrekidis}.} \bibinfo{year}{2012}\natexlab{}.
\newblock \showarticletitle{Accelerating agent-based computation of complex
  urban systems}.
\newblock \bibinfo{journal}{\emph{International Journal of Geographical
  Information Science}} \bibinfo{volume}{26}, \bibinfo{number}{10}
  (\bibinfo{year}{2012}), \bibinfo{pages}{1917--1937}.
\newblock


\end{thebibliography}
	
	%-------------------------------------------------------------------------
	%-------------------------------------------------------------------------
	%-------------------------------------------------------------------------
	%-------------------------------------------------------------------------
	
	\begin{appendices}
		\section{Appendix}
		\subsection{Extensions to CGA building generation}\label{CGA_extensions}
		
		Two new operations have been added to the CGA language in order to add semantic information to the generated shapes. These operations can be used in the successors of the user-written rules, and affect the current \textit{CGA scope} at that point (i.e. the ``bounding box''):
		\begin{itemize}
			\item \texttt{entrance("btype")}, \textit{btype $\in$ \{school, house, shop, workplace, leisure, etc\}}: creates an entrance to a building of the user-defined type, at the origin of the current scope.
			\item \texttt{zone("ztype")}, \textit{ztype $\in$ \{park, road, bike lane, etc\}}: defines a zone of interest of an user-defined type spanning the current scope. If it is a 2D scope (e.g. polygon), it will be extended into a prism along its normal direction.
		\end{itemize}
		A new annotation has also been included, which can be written right before a rule:
		\begin{itemize}
			\item \texttt{@Object("otype")}, \textit{otype $\in$ \{bench, light, fountain, etc\}}: tags the rule production and all its sub-productions as a separate entity to the lot geometry, and marks it with the specified user-defined tag.
		\end{itemize}
		
		%-------------------------------------------------------------------------
		%\subsection*{\emph{PCG} syntax}\label{PCG_syntax}
		\subsection{PCG syntax}\label{PCG_syntax}
		
		An agenda rule set starts by a declaration section where typically a set of attributes (whose values may be overridden by the execution input) and constant values are initialized with the simple syntax:
		
		\begin{center}
			\texttt{variable = ${<}value{>}$}
		\end{center}
		
		The body of the rule set is composed by a sequence of \textbf{rules}, each one of the form:
		
		\begin{center}
			\texttt{predecessor -> successor}
		\end{center}
		
		Where the \texttt{predecessor} is simply an identifier label, and the \texttt{successor} consists of an ordered sequence of parameterized commands (which resemble function calls). Each command may be a recursive call to execute another rule, or the call to an \textbf{operation} which as side-effect alters the current execution context. The arguments passed as parameters may be constant values, variables, calls to \textbf{functions} expressions combining them (using values and/or variables).
		
		The \texttt{@StartRule} annotation before a rule predecessor is used to mark the default starting rule of the rule set itself. During an agenda generation, the commands in the successor of a rule are executed \textbf{sequentially}, performing the rule and operation calls.
		
		% Esto no cuadra
		%These rules are assigned at initialization time to each agenda, and they are iteratively checked for each agent during execution, allowing not only the possibility of dynamic conditions, but also the possibility of dynamically editing the rule set on the fly.
		
		Similarly to CGA, our PCG defines a set of global read-only variables, as well as a set of supported functions (which return some value and can be used inside expressions) and operations (which can be used in the body of the rules producing side-effects to the execution context). The subsequent appendices describe respectively the supported sets.
		
		%-------------------------------------------------------------------------
		%\subsection*{Global variables in \emph{PCG}}\label{PCG_globals}
		\subsection{Global variables in PCG}\label{PCG_globals}
		
		Some built-in read-only global variables exist, which can be accessed to query attributes of the household or of the currently focused person. Currently PCG supports the following variables:
		
		\begin{itemize}
			\item \textbf{home}: The ID of the building where the household lives.
			\item \textbf{gender}: A Boolean which represents the person gender (True for woman and False for man).
			\item \textbf{age}: A number representing the person age.
			\item \textbf{person.id}: ID of the currently focused person (-1 if none).
			\item \textbf{household.id}: The ID of the current household.
		\end{itemize}
		
		%-------------------------------------------------------------------------
		%\subsection*{Functions in \emph{PCG}}\label{PCG_functions}
		\subsection{Functions in PCG}\label{PCG_functions}
		
		Functions in PCG are used to encapsulate calculations, or query information about the simulation world (e.g. city properties), and are not used directly in rules, but they return a value and are used expressions as parameters for other functions or operations. Some functions expect parameters which are numerical references to buildings, citizens or households. The system provides some built-in functions, but users may define custom ones. The current list of built-in functions in \emph{PCG} is:
		
		\begin{itemize}
			\item \textbf{getDistance(}\textit{${<}building1{>},{<}building2{>}$}\textbf{)}: Computes the distance on the navmesh between entrances of two given buildings.
			\item \textbf{getDistanceInTime(}\textit{${<}building1{>}$,${<}building2{>}$}\textbf{)}: Computes the expected time required to travel between the two specified building entrances. Requires having a person focused, as it uses the person's walk speed to perform the computation.
			\item \textbf{findBuilding(}\textit{${<}buildingType{>}$}\textbf{)}: Returns a reference to a building of the given type, selected at random.
			\item \textbf{findNearestBuilding(}{${<}buildingType{>},{<}buildingID{>}$}\textbf{)}: Returns a reference to the building of the given type closest to the building provided as last parameter.
			\item \textbf{findObject(}\textit{${<}type{>},{<}radius{>}$}\textbf{)}: Returns a reference to a random object of the given type, like benches, parks, or any other object within the city. If a radius is provided, only objects within that distance from the current person are considered.
			\item \textbf{isValid(}\textit{${<}obj{>}$}\textbf{)}: Returns a Boolean indicating whether a returned building, object or zone ID is valid or not. Can be used to check whether the functions to find items failed or not, and write fallback behaviors.
			\item \textbf{count(}{${<}predicate{>}$}\textbf{)}: counts the number of household members that match the specified criteria. If no predicate is given, returns the total number of members.
			\item \textbf{chooseMember(}\textit{${<}predicate{>}$}\textbf{)}: function that chooses and returns the ID of a member of the household among the ones matching the specified criteria. This function can be used, for example, to choose an adult from the household that can carry out a required action. The predicate may be also an heuristic scoring function, in which case the member that is evaluated with the largest value will be selected (or if multiple ones are, one of them is randomly chosen.)
		\end{itemize}
		
		%-------------------------------------------------------------------------
		\subsection{Operations} \label{PCG_operations}
		Operations can be used in rules, and as opposed to the functions, they do not return a value. Instead, they cause some effect on the current execution context such as creating a temporary variable, or creating new agenda tasks. \\
		
		The main set of the supported operations operations is aimed at inserting different new actions into the agenda, and most of them take as arguments a start (\textit{$t_0$}) and end (\textit{$t_1$}) times. For these operations to have an effect, a member of the household must be focused. Then this focused member will be the only one affected. These operations are:
		
		\begin{itemize}
			\item \textbf{stayInside(}\textit{${<}t_0{>},{<}t_1{>},{<}buildingID{>}$}\textbf{)}: instructs the person to stay ``hidden'' inside the given building between times $t_0$ and $t_1$.
			\item \textbf{goToBuilding(}\textit{${<}t_0{>},{<}t_1{>},{<}buildingID{>}$}\textbf{)}: makes the person walk its way to the specified building starting at time $t_0$, and enter it when reached, remaining inside until time $t_1$. The difference with the previous one is where they are expected to be at the initial time.
			\item \textbf{accompany(}\textit{${<}time{>},{<}condition{>}$}\textbf{)}: creates a shared group task where the current person is the "leader", and other household members matching the condition join. The task goal and start/end times are copied from the first non-leader member agenda, taking the agenda entry that would execute at the time specified as parameter. This operation is used in our examples to create tasks for a parent to accompany their children to school.
			\item \textbf{delayedRule(}\textit{${<}t_0{>},{<}t_1{>},{<}ruleName{>}$}\textbf{)}: creates a delayed-evaluation order for the person to execute a rule as a task in a specified time slot. The specified rule will be used as the starting point to dynamically generate a behavior for the person. This takes into account factors such as resource availability (e.g. occupied park benches) and creates the basis for other emergent behaviors. The delayed rule executions have some particularities (detailed in the next section), and a different set of possible rule operations, as listed below.
			\item \textbf{floatingSlot(}\textit{${<}t_0{>},{<}t_1{>}$}\textbf{)}: Specifies that the user has "free time" in a period, and can execute one or multiple tasks from the pool of floating tasks as long as they fit in this slot starting and ending time. When there are no suitable tasks, the person defaults to remain inside their house.
		\end{itemize}
		
		This other set of operations can pause the execution until a given condition is meet, for instance after a certain time has passed, and are used in the rules triggered in delayed execution:
		\begin{itemize}
			\item \textbf{wait(}\textit{${<}seconds{>}$}\textbf{)}: keeps the person idle standing still for the specified duration.
			\item \textbf{goToZone(}\textit{${<}zoneType{>}$}\textbf{)}: finds and entrance to the nearest zone of the given type, and makes the person walk to it, blocking the execution until it is reached.
			\item \textbf{goToObject(}\textit{${<}objectId{>}$}\textbf{)}: makes the person walk to the given object, and stalls execution until it is reached.
			\item \textbf{interact(}\textit{${<}objectId{>}$}\textbf{)}: interacts with the specified object (if that is possible). For instance, in the case of a bench it means to sit there.
		\end{itemize}
		
		Finally, a few other operations have unique effects:
		
		\begin{itemize}
			
			\item \textbf{members \{}\textit{ ${<}cond1{>}$}\textbf{:} \textit{${<}actions1{>}$} \textbf{$|$} \textit{${<}cond2{>}$}\textbf{:} \textit{${<}actions2{>}$} \textbf{$|$} ...\textbf{\}}: for each member in a household, each condition is verified in order and -only- the matching entry, if any, has its actions executed.
			
			\item \textbf{floatingTask(}\textit{${<}duration{>},{<}ruleName{>},[{<}priority{>}]$}\textbf{)}: Adds a new task to the pool of floating tasks, to execute the specified rule (as a delayed rule execution) with the specified maximum duration. The priority is an optional real number, which defaults to 0.
			
		\end{itemize}

		\subsection{PCG Example}\label{PCG_example}
		Here we can see an example of the rule file for an agenda generation
		%\begin{minipage}{\textwidth}
		\vbox{
			\footnotesize
			\begin{lstlisting}[frame=single,style=CGA,caption=Example of an agenda generation rule file,label=code:AgendaRuleFileExample, basicstyle=\small] %,float=*]
			
			schoolStart = 8h    schoolEnd = 16h
			workStart   = 8h    workEnd   = 16h
			
			@StartRule
			Household -->
			members{true: stayInside(0h,24h,home)} 
			members { age < 18: ChildrenWeekday
			| age < 65: AdultWeekday
			| true:     ElderWeekday }
			
			case count(age < 18) != 0:
			members { chooseMember(age >= 18):
			BringChildrenToSchool}
			
			VisitBuilding(t0,t1,building) -->
			set(time, getDistanceInTime(home, building))
			goToBuilding(t0 - time, t0, building)
			stayInside(t0, t1, building)
			goToBuilding(t1, t1 + time, home)
			
			ChildrenWeekday -->
			set(school,findNearestBuilding("school",home))
			VisitBuilding(schoolStart, schoolEnd, school)
			
			AdultWeekday -->
			set(workplace, findBuilding("workplace"))
			VisitBuilding(workStart + rand(-5m, 5m),
			workEnd, workplace)
			
			ElderWeekday --> 
			delayedRule(8h + rand(-30m, 30m), 12h + rand(-30m, 30m), VisitPark)
			
			BringChildrenToSchool -->
			accompany(schoolStart-2, age < 18)
			
			GoToPark -->
			wait(2)
			goToZone("park")
			SitInBench
			
			SitInBench -->
			set(bench, findObject("bench", 20))
			[
			case isValid(bench):
			goToObject(bench)
			interact(bench, "sit")
			waitUntilNextTask()
			else:
			...
			]
			
			\end{lstlisting}
			\normalsize
		}
		%\end{minipage}
		
	\end{appendices}
	%-------------------------------------------------------------------------
	%-------------------------------------------------------------------------
	%-------------------------------------------------------------------------
	%-------------------------------------------------------------------------
	%-------------------------------------------------------------------------
\end{document}